\documentclass[twocolumn,trackchanges]{aastex701}
\usepackage{rotating}
\begin{document}

\title{A Search for Variability of Ultracool Dwarfs with the Zwicky Transient Facility}

\correspondingauthor{Shu Wang}
\email{shuwang@nao.cas.cn}

\author[0009-0004-3407-0848]{Haomiao Huang}
\affiliation{CAS Key Laboratory of Optical Astronomy, National Astronomical Observatories, Chinese Academy of Sciences, Beijing 100101, People's Republic of China}
\affiliation{School of Astronomy and Space Sciences, University of Chinese Academy of Sciences, Beijing 100049, People's Republic of China}
\email{huanghm@bao.ac.cn}  

\author[0000-0003-4489-9794]{Shu Wang}
\affiliation{CAS Key Laboratory of Optical Astronomy, National Astronomical Observatories, Chinese Academy of Sciences, Beijing 100101, People's Republic of China}
\affiliation{School of Astronomy and Space Sciences, University of Chinese Academy of Sciences, Beijing 100049, People's Republic of China}
\email[]{shuwang@nao.cas.cn}

\author[0000-0001-7084-0484]{Xiaodian Chen}
\affiliation{CAS Key Laboratory of Optical Astronomy, National Astronomical Observatories, Chinese Academy of Sciences, Beijing 100101, People's Republic of China}
\affiliation{School of Astronomy and Space Sciences, University of Chinese Academy of Sciences, Beijing 100049, People's Republic of China}
\affiliation{Institute for Frontiers in Astronomy and Astrophysics, Beijing Normal University, Beijing 102206, People's Republic of China}
\email{}

\author[0009-0000-7976-7383]{Zhijun Tu}
\affiliation{CAS Key Laboratory of Optical Astronomy, National Astronomical Observatories, Chinese Academy of Sciences, Beijing 100101, People's Republic of China}
\email{zjtu@bao.ac.cn}

\author{Jifeng Liu}
\affiliation{CAS Key Laboratory of Optical Astronomy, National Astronomical Observatories, Chinese Academy of Sciences, Beijing 100101, People's Republic of China}
\affiliation{School of Astronomy and Space Sciences, University of Chinese Academy of Sciences, Beijing 100049, People's Republic of China}
\affiliation{Institute for Frontiers in Astronomy and Astrophysics, Beijing Normal University, Beijing 102206, People's Republic of China}
\affiliation{New Cornerstone Science Laboratory, National Astronomical Observatories, Chinese Academy of Sciences, Beijing 100012, People's Republic of China}
\email{jfliu@nao.cas.cn}

\begin{abstract}

Rotationally modulated photometric variability of ultracool dwarfs encodes key information about cloud structure and temperature contrasts. Large homogeneous optical datasets are crucial for linking atmospheric heterogeneity to fundamental parameters such as rotation, mass, and age. We present a search for rotation periods in ultracool dwarfs using Zwicky Transient Facility (ZTF) optical light curves. By propagating the coordinates to the ZTF epoch and applying Lomb-Scargle analysis, we identified 226 periodic variables, including 32 robust detections and 194 tentative cases. Among the robust detections, 12 have no previously published periods, while 20 have literature counterparts, most of which are consistent with the published values. Most robust detections are M dwarfs, reflecting the optical sensitivity limits of ZTF. We find a trend of decreasing periods toward later spectral types in relatively old dwarfs ($>$100 Myr), suggesting faster rotation for late-M types than for mid-M types. The age-period relation of our sample is broadly consistent with angular-momentum-conservation models at higher-mass regime of brown dwarfs, consistent with the M-dwarf bias of our catalog. Many additional candidates remain to be confirmed due to sparse sampling or low S/N. Future high-cadence, multi-wavelength monitoring and systematic mining of ZTF and upcoming surveys will be crucial for validating these periods, extend sensitivity to later (L/T) types, and better connect rotation with cloud physics across the stellar–substellar boundary.

\end{abstract}



\section{Introduction} 

Brown dwarfs are substellar objects with masses ranging from the heaviest gas giant planets to the lightest stars \citep[$\sim$13--80 $M_{\rm Jup}$,][]{2001RvMP...73..719B}, which are insufficient to sustain stable hydrogen fusion in their cores \citep{1963ApJ...137.1121K}. Rotation-modulated photometric variability in brown dwarfs is widely interpreted as the changing visibility of heterogeneous photospheric features, primarily patchy condensate clouds, temperature perturbations, or spots, that rotate in and out of view \citep{2012ApJ...760L..31B, 2015ApJ...799..154M, 2017Sci...357..683A}. Time-resolved photometry and spectroscopy have shown that these surface inhomogeneities produce periodic (and often evolving) light curves \citep{2009ApJ...701.1534A, 2017MNRAS.471..811B}. The amplitudes of the variability and its wavelength and spectral type dependence encode information on vertical cloud structure, temperature contrasts, and atmospheric dynamics \citep{2012ApJ...750..105R, 2014ApJ...793...75R, 2015ApJ...798L..13Y}. Therefore, studying rotationally modulated photometric variability of ultracool dwarfs provides unique insight into the atmospheric structure and cloud properties of these objects. 

Large, targeted monitoring campaigns and high-precision space- and ground-based programs have mapped out the phenomenology of variability of ultracool dwarfs. Multi-wavelength \textit{Spitzer} Space Telescope surveys and \textit{Hubble} Space Telescope (\textit{HST}) observations have shown that variability is common across L and T dwarfs, with especially large amplitudes near the L/T transition\citep{2013ApJ...768..121A, 2015ApJ...799..154M}, and that their rotation periods span from $\sim$1.4 hr to tens of hours \citep{2008MNRAS.386.2009C,2015ApJ...799..154M}. For late-type M dwarfs, extensive photometric monitoring has revealed that rotation periods span from a few hours to several tens of days, with shorter periods generally corresponding to higher photometric amplitudes \citep{2016ApJ...821...93N, 2011ApJ...727...56I, 2014ApJS..211...24M}. The light-curve morphologies of ultracool dwarfs are diverse, ranging from stable single-period signals to multi-component or rapidly evolving behavior \citep{2009ApJ...701.1534A, 2017Sci...357..683A}, and viewing geometry and rotation rate modulate the observed signals \citep{2017Sci...357..683A, 2017ApJ...842...78V}. Together with Doppler imaging and spectroscopic $v \sin i$ studies, these results establish that brown dwarfs are rapid rotators with diverse weather patterns and evolutionary timescales, making them excellent laboratories for comparative cloud physics \citep{2014Natur.505..654C, 2014A&A...566A.130C, 2008ApJ...684.1390R}.

The fundamental parameters of brown dwarfs like mass and radius, are difficult to measure directly. Brown-dwarf eclipsing binaries provide the most direct constraints on these properties. The first such system, 2MASS J05352184$-$0546085 in the Orion Nebula Cluster \citep{2006Natur.440..311S}, showed that young brown dwarfs can have radii comparable to those of low-mass stars. The 45 Myr-old system 2MASS J1510478$-$281817 \citep{2020NatAs...4..650T} provides a later-stage benchmark. Such systems remain important for calibrating substellar evolutionary models.

The Zwicky Transient Facility (ZTF) offers an attractive complement to previous variability programs because of its very large areal coverage and long-duration optical monitoring \citep{2019PASP..131f8003B, 2020ApJS..249...18C}. To date, the literature contains rotation periods for on the order of $\sim$200 ultracool dwarfs at spectral types $\ge$M5 \citep[e.g.][]{2013MNRAS.432.1203M,2016ApJ...821...93N,2021AJ....161..224T,2023MNRAS.521..952M,2024MNRAS.527.8290P}, while the published sample in the late-M to early-L range most accessible to optical surveys remains more limited, comprising only 94 objects in the M5–L0 interval. ZTF is particularly powerful for discovering and characterizing periodic variability in relatively bright, nearby ultracool objects, and for assembling large, homogeneous catalogs of optical periods and long-term evolution. Although ZTF’s optical bandpasses and typical photometric precision are less sensitive to intrinsically faint late-L, T, and Y dwarfs and to the smallest-amplitude variations revealed by space-based infrared programs \citep{2015ApJ...799..154M, 2012ApJ...760L..31B}, and the sparse photometric cadences of ZTF also make it difficult to determine the periods of sources whose atmospheric conditions change in short timescales \citep{2009ApJ...701.1534A}, its extensive baseline and large sample size provide a valuable opportunity to extract reliable periods. By applying strict quality cuts, splicing same-filter light curves, using multi-band information as auxiliary support, and adopting statistical criteria, we mitigate the challenges posed by sparse or low-S/N data. These efforts make ZTF a powerful tool for investigating brown dwarf variability, despite its inherent limitations.

In this paper, we exploit ZTF photometry to build a catalog of periodic ultracool variables. We identify 32 robust periodic variables, including 12 no previously published periods, increasing the published $\ge$M5 period sample by $\sim$6\% (see Section~\ref{sec:5.3}). We organize the paper as follows. In Section \ref{sec:2}, we provide a concise introduction to the ZTF survey and specify the spectral type of ultracool dwarfs for which ZTF can be used to carry out variability studies. In Section \ref{sec:3}, we describe our sample selection and data inspection. Section \ref{sec:4} presents the period detection. Section \ref{sec:5} presents the resulting catalog, period accuracy tests, and newly identified variables. In Section \ref{sec:6}, we discuss the relations between period, amplitude, spectral type and age. Section \ref{sec:7} summarizes our conclusions.

\section{ZTF data} \label{sec:2}

\begin{figure*}[ht!]
\includegraphics[width=1\textwidth]{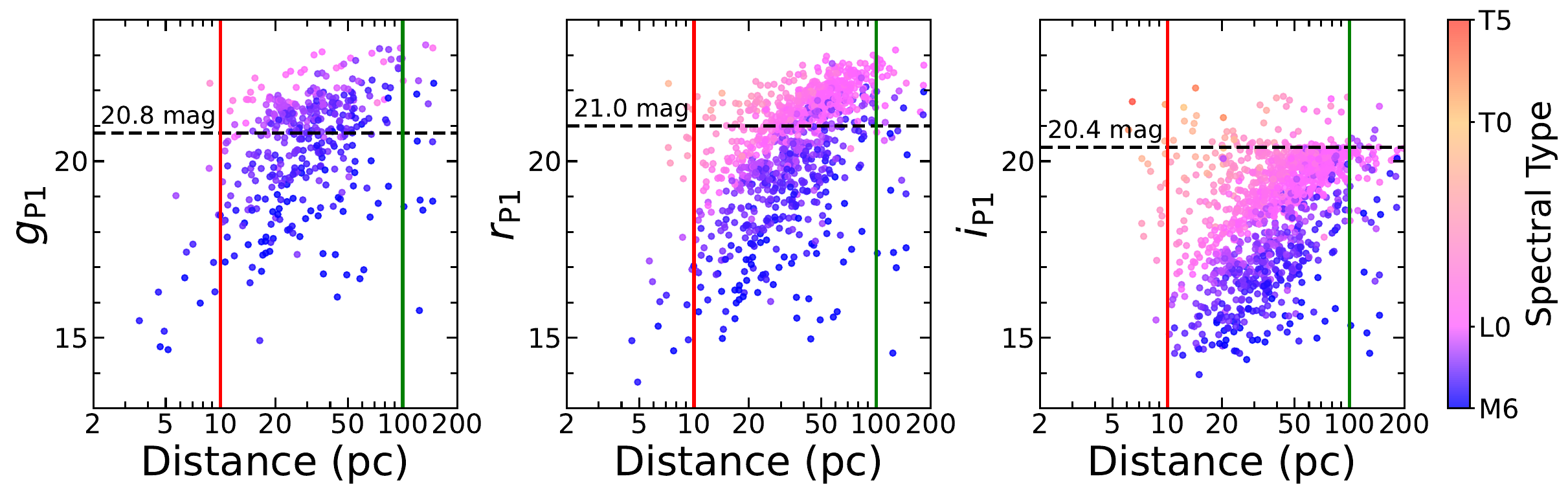}
\caption{PS1 $g$, $r$ and $i$ magnitudes of ultracool dwarfs plotted against distance, with points colored by spectral type (see color bar). Red and green vertical lines mark distances of 10 pc and 100 pc, respectively. The black dashed lines indicate the PS1 magnitudes corresponding to the ZTF detection limits, estimated from the mean magnitude offsets between PS1 and ZTF photometric systems together with the ZTF limiting magnitudes (see text for details). In the $g$, $r$ and $i$ bands at 10 pc, only sources of spectral types not later than L0, L5 and L8, respectively, are detectable; at 100 pc, only sources of spectral types not later than M6, L2, and L3, respectively, are detectable. The PS1--ZTF offsets and the implied boundaries are approximate and are shown for illustrative purposes only. \label{fig:1}}
\end{figure*}

ZTF is an optical time-domain survey conducted using the 48‑inch Schmidt telescope at Palomar Observatory \citep{2019PASP..131f8003B,2019PASP..131a8003M}. Equipped with a custom-built wide-field camera, it offers an expansive 47 deg$^2$ field of view and 8 second readout time, making ZTF a powerful facility for detecting transient objects and stellar variability \citep{2019PASP..131g8001G}. Starting from its commissioning in 2018 March 17, the public observing time was allocated as follows: during Phase I (March 2018 $-$ September 2020), 40\% was dedicated to public surveys; in Phase II (from December 2020 onward), this increased to 50\%. These public observations focused primarily on two surveys: the Northern-equatorial sky survey and the Galactic plane-targeted survey. The Northern hemisphere ($\sim$27,500 deg$^2$) is revisited every three nights in Phase I and every two nights in Phase II, while the Galactic plane region ($\left | b \right | \le 7^{\circ}$, $\delta>\sim-25^{\circ}$) is monitored nightly. Observations are carried out in $g$-, $r$-, and $i$-bands, with typical limiting magnitudes of $g\sim$ 20.8 mag, $r\sim$ 20.6 mag and $i\sim$ 19.9 mag at 5$\sigma$ in 30 second exposures. We utilized ZTF DR23, which comprises public-sky data from March 2018 to October 2024, and private-survey data obtained from March 2018 to June 2023.

The pioneering work by \citet{2020ApJS..249...18C} systematically searched for variability using ZTF DR2 and constructed a large-scale catalog of periodic variable stars. Our study focuses on variability in ultracool dwarfs. The long-term photometric monitoring of ZTF provides an extended temporal baseline, addressing the current shortage of prolonged coverage in studies of ultracool star variability \citep[e.g.][the time spans of light curves in these works are often only a few hours to a few days]{2015ApJ...799..154M, 2021AJ....161..224T, 2025A&A...697L..10M, 2020AJ....159...60G}. However, limitations arise when using ZTF data to search periods of ultracool dwarfs. First, as a ground‑based facility, ZTF observations are affected by atmospheric conditions, the diurnal cycle, and seasonal gaps, which limit photometric precision compared to space‑based instruments and introduce aliased periods when running a Lomb–Scargle periodogram \citep{1976Ap&SS..39..447L, 1982ApJ...263..835S} analysis. Second, the typical cadence of ZTF is longer than typical brown-dwarf rotation periods, which typically occur over timescales of hours to days. Therefore, obtaining multiple measurements within a single period is difficult, which makes it difficult to determine the rotation periods of objects with evolving weather patterns that alter their light curves \citep[e.g.][]{2009ApJ...701.1534A, 2013A&A...555L...5G}.

\begin{figure*}[ht!]
\includegraphics[width=1\textwidth]{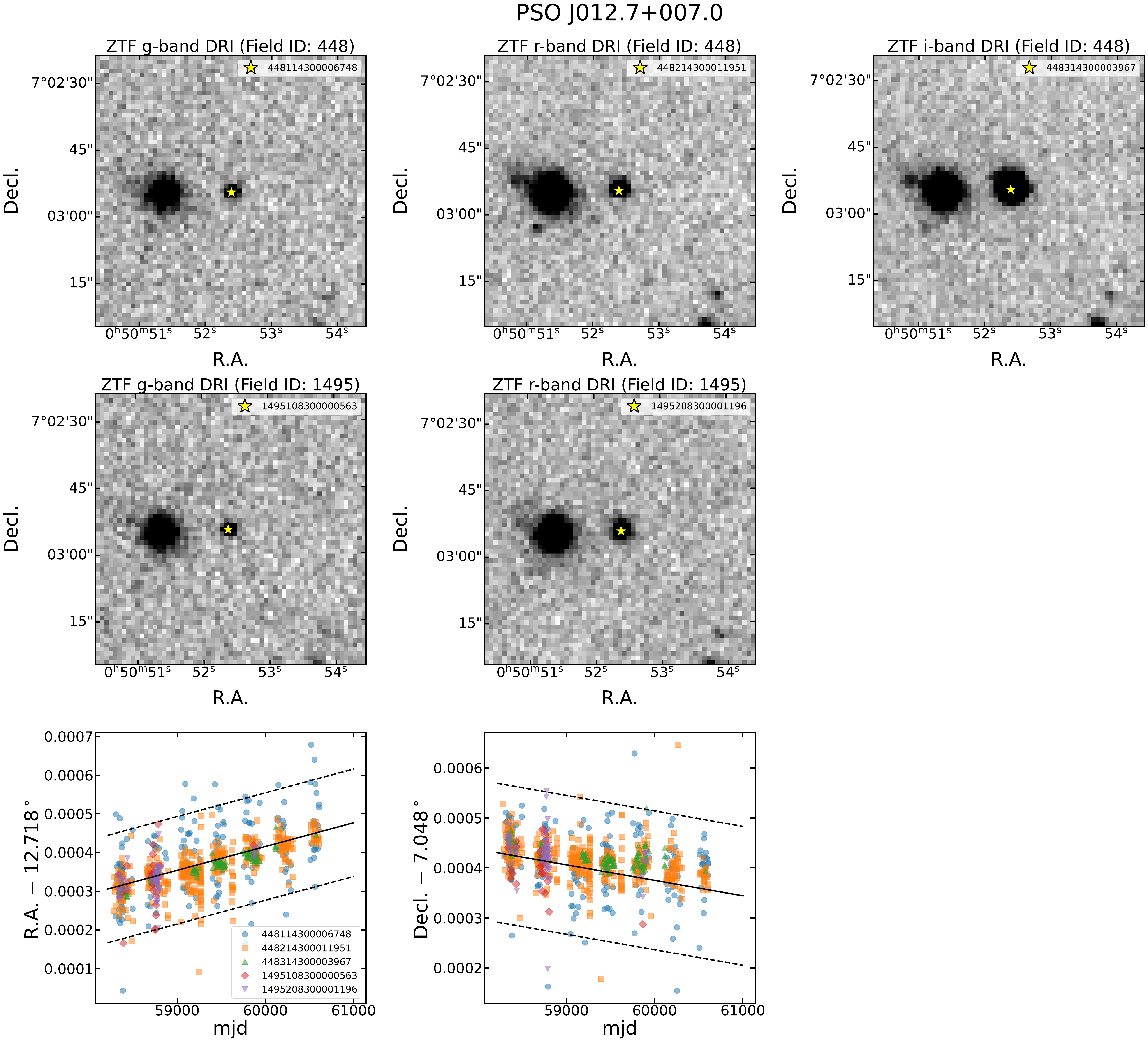}
\caption{Example of light‑curve verification. Top (Field 448) and middle (Field 1495) panels illustrate the coordinate–source coincidence check for the brown dwarf PSO\,J012.7+007.0 (M7-type), with light‑curve coordinates of different internal product ID marked by yellow star symbols overlaid on ZTF $g$‑, $r$‑, and $i$‑band Deep Reference Images (DRI) from left to right. The increasing brightness from $g$- to $i$-bands confirms the object’s low temperature nature. The bottom panel shows the coordinate shifts compared to the expected proper‑motion. The solid black line represents predicted change in coordinates over time derived from the Gaia EDR3 position and proper motion, which is similar to the observed coordinate variation of this source in ZTF. The dashed black lines on both sides deviate from the solid line by 0$\farcs$5. The legends indicate different internal product IDs for this source in ZTF, with distinct colors and markers. The complete figure set (32 images) is available in the online journal. \label{fig:2}}
\end{figure*}

The peak emission of ultracool dwarfs does not lie in the optical band, and their intrinsically low luminosities mean that only relatively nearby sources can be detected with sufficient precision for variability studies. Figure \ref{fig:1} presents the PAN‑STARRS1 \citep[PS1,][]{2016arXiv161205560C} $g$, $r$, and $i$ magnitudes as a function of distance for ultracool dwarfs, with different colors denoting different spectral types. PS1 provides a useful reference for this purpose because it offers homogeneous optical photometry over a large sky area and reaches a depth suitable for evaluating the approximate ZTF detectability regime of ultracool dwarfs. It therefore gives an order-of-magnitude view of which spectral types are likely to remain accessible to ZTF at different distances. Since the ZTF filters are custom and their central wavelengths differ from those of PS1, we computed the mean magnitude for each light curve and compared it to the corresponding PS1 magnitude in the nominally corresponding band, finding that ZTF magnitudes are, on average, smaller than PS1 magnitudes by $\sim$0.05 mag in the $g$ (negligible, omitted in Figure \ref{fig:1}), $\sim$0.4 mag in the $r$, and $\sim$ 0.4 mag in the $i$ band. We add these offsets to the limiting magnitudes of the respective ZTF bands to define approximate ZTF detectability boundaries. These PS1--ZTF offsets and the resulting boundaries are intended only to illustrate the strong optical selection effects of ZTF, rather than to provide a precise completeness limit. From Figure \ref{fig:1}, we can roughly estimate that within 10 pc (as indicated by the red vertical line), the $g$, $r$, and $i$ bands can detect sources of spectral types not later than L0, L5, and L8, respectively, whereas within 100 pc (green vertical line) these bands are sensitive only to sources not later than M6 ($g$), L2 ($r$), and L3 ($i$), respectively. These approximate detectability boundaries illustrate the selection effects on the sample’s spectral types and physical properties.

\section{Sample selection and data inspection} \label{sec:3}

\begin{figure*}[ht!]
\includegraphics[width=1\textwidth]{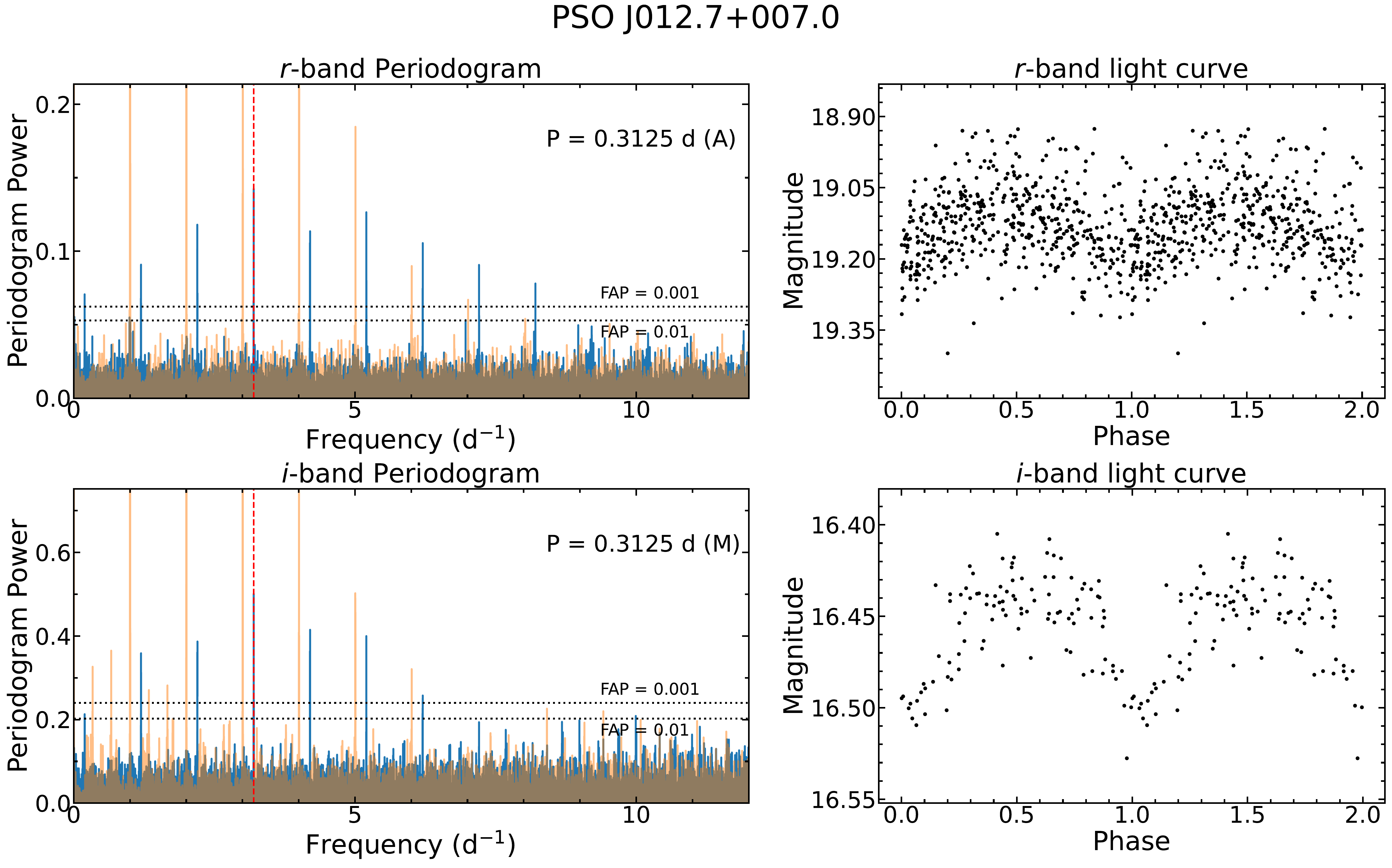}
\caption{Example light-curve analysis of the brown dwarf PSO J012.7+007.0 in the $r$ (top) and $i$ (bottom) bands. The left panels present the periodograms. The blue lines are the periodograms derived from the ZTF light curves, the orange curves indicate the window functions, and the red dashed lines mark the locations of the peaks. The period values labeled in the panels are followed by remark codes in parentheses, indicating whether the period is adopted, matched, or otherwise classified, following the notation used in Appendix~\ref{sec:appendA} and Table~\ref{tab:a1}. The right panels display the phase-folded light curves. No periodicity is detected in the $g$-band light curve of this target due to its faintness in this band. The complete figure set (32 images) is available in the online journal. \label{fig:3}}
\end{figure*}

We employed the UltracoolSheet\footnote{http://bit.ly/UltracoolSheet} \citep{best_2025_15802304} to select suitable targets from the main table. We selected sources whose coordinates lie within the sky coverage of the ZTF survey and required their PS1 $i$‑band magnitudes to be brighter than 20.5 mag. This $i$-band threshold is a practical pre-selection for retrieving and inspecting light curves, rather than a strict ZTF detection limit, and was chosen to allow for scatter in the approximate PS1--ZTF photometric transformation while maintaining a reasonable likelihood of useful ZTF photometry for period searches. For sources lacking PS1 $i$-band photometry, we estimated PS1 $i$ magnitudes from empirical relations calibrated with objects having both PS1 $i$-band and infrared measurements, including the Two Micron All-Sky Survey (2MASS; $J$, $H$, $K_s$; \citealt{2006AJ....131.1163S}) and the Wide-field Infrared Survey Explorer (WISE; $W1$, $W2$; \citealt{2010AJ....140.1868W}) photometry, and adopting the priority $J>H>K_s>W1>W2$ because the $J$ band lies closer to the optical regime. Because T- and Y-type brown dwarfs with measured $r$- and $i$-band magnitudes are scarce, our magnitude relations are based predominantly on M and early‑L dwarfs, which may introduce systematic errors in the estimated magnitudes of T‑ and Y‑type objects. However, given the approximate detectability boundaries shown in Figure \ref{fig:1}, T‑ and Y‑type brown dwarfs are too faint to be detectable by ZTF, making these systematic uncertainties negligible for our study.

Furthermore, Figure \ref{fig:1} indicates that the ultracool dwarfs detectable by ZTF are generally relatively nearby, implying they usually exhibit large proper motions, up to several hundred to a few thousand mas yr$^{-1}$. If one simply searches using coordinates at the epoch of J2000, the true counterparts often cannot be recovered. To retrieve the correct light curves, we propagated the source positions to the survey epoch of March 2018, giving priority to Gaia coordinates at Ep $= 2016.0$ and proper motions \citep{2021A&A...649A...1G,2023A&A...674A...1G} for coordinate transformation. The resulting epoch‑2018 coordinates were then used to query ZTF. We downloaded all light curves within a 2$^{\prime\prime}$ search radius and containing at least 20 detections from the NASA/IPAC Infrared Science Archive (IRSA)\footnote{https://irsa.ipac.caltech.edu/Missions/ztf.html}.

To avoid erroneous downloads caused by incorrect transformed coordinates derived from potentially inaccurate proper motions, we performed a two-part verification for each retrieved light curve. First, we overplotted the epoch‑2018 coordinates on deep reference images to confirm visually that a source is present at that location, and that its brightnesses in the $g$-, $r$-, and $i$-band images increase as expected for ultracool dwarfs. Second, we checked that the R.A. and Decl. values of the light curve closely match the epoch-2018 coordinates and that their temporal shifts are broadly consistent with the proper motion from the literature, while allowing for modest discrepancies between the observed and expected coordinate shifts due to uncertainties in the proper motions. Only sources satisfying both criteria were retained for variability analysis. Starting from 3890 objects in the UltracoolSheet main table, 2004 satisfied the ZTF sky-coverage and brightness pre-selection criteria described above. Of these, 1961 sources were successfully propagated to the 2018 epoch and queried in ZTF, and 1538 yielded at least one retrieved light curve. The subsequent visual/astrometric verification and the requirement of sufficient usable photometric points were applied (Section~\ref{sec:4}) jointly in our workflow rather than as separately tabulated stages. After these checks, 1492 sources remained for period analysis. Figure \ref{fig:2} provides an example of this verification.

\section{Detection of Periodic Variables} \label{sec:4}

\begin{deluxetable*}{hlDDcDDDcchlcrcl}

\tabletypesize{\scriptsize}

\tablecaption{Catalog of periodic ultracool variables detected in ZTF}\label{tab:1}

\tablenum{1}

\tablehead{\nocolhead{Shortname} & \colhead{Name} & \multicolumn2c{R.A. (J2000)\tablenotemark{a}} & \multicolumn2c{Decl. (J2000)\tablenotemark{a}} & \colhead{Band} & \multicolumn2c{Magnitude} & \multicolumn2c{Period} & \multicolumn2c{log(FAP)} & \colhead{Amplitude} & \colhead{R$^2$} & \nocolhead{N} & \colhead{SpT} & \colhead{P$_{lit}$} & \colhead{Ref.} & \colhead{Reliability\tablenotemark{b}} & \colhead{Multi.\tablenotemark{c}} \\
\nocolhead{} & \colhead{} & \multicolumn2c{(deg)} & \multicolumn2c{(deg)} & \colhead{} & \multicolumn2c{(mag)} & \multicolumn2c{(days)} & \multicolumn2c{} & \colhead{(mag)} & \colhead{} & \nocolhead{} & \colhead{} & \colhead{(days)} & \colhead{} & \colhead{} & \colhead{} }
\decimals
\startdata
 & WISEA J000538.83+020951.2 & 1.4111 & 2.1651 & r & 16.731 & 0.9461404 & -23.295 & 0.0461 & 0.2536 & 430 & M6 &  &  & R & Y* \\
J0027+2239B & LP 349-25 & 6.9834 & 22.3257 & r & 16.779 & 0.0841378 & -5.758 & 0.0218 & 0.0809 & 557 & M8 & 0.0775 & 1 & R &  \\
 & PSO J012.7+007.0 & 12.7179 & 7.0486 & r & 19.130 & 0.3125428 & -12.081 & 0.0938 & 0.1425 & 481 & M7.1 &  &  & R & Y* \\
J0129-0823 & 2MASS J01294256-0823580 & 22.4274 & -8.3995 & r & 15.237 & 0.2655820 & -5.378 & 0.0213 & 0.1217 & 360 & M5 & 0.3115 & 2 & R &  \\
J0217+3526 & LP 245-10 & 34.2917 & 35.4424 & r & 14.912 & 0.2758886 & -13.053 & 0.0285 & 0.1598 & 490 & M5 & 0.276 & 3 & R & Y* \\
J0218-0617 & 2MASS J02185792-0617499 & 34.7412 & -6.2972 & i & 16.011 & 0.0466091 & -4.162 & 0.0394 & 0.3552 & 80 & M8 &  &  & R &  \\
J0235-0711 & DENIS J023549.5-071121 & 38.9566 & -7.1892 & r & 17.716 & 0.3248738 & -11.472 & 0.0369 & 0.1109 & 625 & M7 &  &  & R & Y* \\
J0311+0106 & 2MASS J03111547+0106307 & 47.8145 & 1.1086 & r & 15.278 & 0.1244036 & -17.031 & 0.0377 & 0.2195 & 418 & M5.5 &  &  & R & Y* \\
J0330+2405 & 2MASS J03300506+2405281 & 52.5212 & 24.0911 & r & 17.617 & 0.1978335 & -33.530 & 0.0650 & 0.1925 & 929 & M7 &  &  & R & Y* \\
J0357+4107 & 2MASS J03571999+4107426 & 59.3332 & 41.1285 & r & 15.986 & 0.5672745 & -100.581 & 0.0466 & 0.2459 & 1844 & M6 & 0.568 & 4 & R & Y* \\
J0421+1929 & LP 415-20 & 65.4566 & 19.4857 & r & 17.974 & 0.2050021 & -13.741 & 0.0524 & 0.1461 & 552 & M7.5 & 0.205 & 5 & R &  \\
J0422+1530 & 2MASS J04221413+1530525 & 65.5589 & 15.5146 & r & 18.698 & 0.9354520 & -14.474 & 0.0980 & 0.1646 & 484 & M6 &  &  & R &  \\
J0510+2714 & LSR J0510+2713 & 77.5837 & 27.2339 & r & 16.549 & 0.0773407 & -4.470 & 0.0384 & 0.1316 & 273 & M8 & 4.297 & 6 & R &  \\
J0526-1824 & 2MASS J05264316-1824315 & 81.6799 & -18.4087 & r & 17.593 & 0.4842631 & -7.883 & 0.0433 & 0.1625 & 335 & M6.2 &  &  & R &  \\
J0752+1612 & LP 423-31 & 118.0997 & 16.2042 & r & 16.041 & 7.7883021 & -34.082 & 0.0290 & 0.3409 & 580 & M7 & 0.884 & 3 & R & Y* \\
J0807+3213 & 2MASS J08072607+3213101 & 121.8586 & 32.2195 & i & 15.502 & 0.3450465 & -4.566 & 0.0511 & 0.488 & 63 & M8 & 0.345 & 7 & R & Y* \\
J0835+1029 & 2MASS J08352366+1029318 & 128.8486 & 10.4922 & i & 16.516 & 0.1906884 & -5.025 & 0.0381 & 0.4853 & 63 & M7 & 0.648 & 8 & R &  \\
J0911+2248 & SDSS J091130.53+224810.7 & 137.8774 & 22.8030 & r & 16.581 & 0.1020390 & -5.028 & 0.0178 & 0.0595 & 714 & M7 &  &  & R &  \\
J1003-0105 & 2MASS J10031918-0105079 & 150.8298 & -1.0856 & r & 18.262 & 0.5893867 & -4.476 & 0.0659 & 0.1357 & 255 & M7 & 0.2127 & 9 & R &  \\
J1314+1319B & LSPM J1314+1320 & 198.5848 & 13.3335 & r & 14.737 & 0.1547113 & -22.495 & 0.0496 & 0.2662 & 402 & M7 & 0.158 & 3 & R & Y* \\
J1438+5559 & 2MASS J14384542+5559134 & 219.6893 & 55.9870 & i & 16.006 & 2.0225841 & -7.828 & 0.0113 & 0.0666 & 790 & M7 &  &  & R & Y* \\
J1444+3002 & LP 326-21 & 221.0716 & 30.0373 & i & 14.896 & 0.3712801 & -4.930 & 0.0242 & 0.2664 & 138 & M8 & 0.3706 & 10 & R &  \\
J1609-2007 & [SCH2006] J16091837-20073523 & 242.3265 & -20.1264 & r & 18.291 & 19.2457228 & -9.257 & 0.1202 & 0.2837 & 189 & M7.5 & 18.9678 & 11 & R &  \\
J1627-2441 & SR12 C & 246.8323 & -24.6970 & g & 13.954 & 3.9257864 & -15.506 & 0.2256 & 0.6069 & 101 & M9 & 3.9176 & 11 & R &  \\
J1640+6736 & G 240-23 & 250.0860 & 67.6013 & r & 14.663 & 0.3782275 & -6.550 & 0.0201 & 0.0725 & 685 & M5.5 & 0.378 & 12 & R & Y \\
J1707+6439 & 2MASS J17071830+6439331 & 256.8263 & 64.6592 & i & 16.033 & 0.1511430 & -8.082 & 0.0197 & 0.1772 & 277 & M9 & 0.1508 & 13 & R &  \\
J1835+3259 & 2MASSI J1835379+325954 & 278.9078 & 32.9981 & r & 16.772 & 0.1183858 & -26.828 & 0.0185 & 0.0869 & 1749 & M8.5 & 0.1183 & 14 & R &  \\
J2202-1109 & 2MASS J22021125-1109461 & 330.5469 & -11.1629 & r & 17.984 & 0.4283621 & -6.591 & 0.0459 & 0.123 & 416 & M6.5 & 0.42 & 15 & R &  \\
J2307+0520 & WISEA J230743.63+052037.3 & 346.9322 & 5.3440 & i & 17.698 & 0.0916088 & -5.273 & 0.0448 & 0.2493 & 141 & M7 &  &  & R & Y \\
J2315-0627 & LP 702-50 & 348.9771 & -6.4629 & r & 16.045 & 0.1454282 & -6.707 & 0.0305 & 0.1483 & 318 & M5.5 &  &  & R &  \\
J2338+0624 & LP 583-21 & 354.6413 & 6.4144 & r & 16.309 & 0.2514961 & -13.266 & 0.1515 & 0.7614 & 69 & M7 & 0.251 & 3 & R & Y* \\
J2353-1844 & 2MASS J23532556-1844402 & 358.3565 & -18.7446 & r & 15.762 & 0.6167348 & -12.825 & 0.0268 & 0.169 & 482 & M5.8 & 0.6166 & 16 & R & Y* \\
 & PSO J000.2794+16.6237 & 0.2792 & 16.6241 & r & 20.699 & 0.0568013 & -3.290 & 0.1594 & 0.0918 & 312 & M7.2 &  &  & C &  \\
$\cdots$ & $\cdots$ & $\cdots$ & $\cdots$ & $\cdots$ & $\cdots$ & $\cdots$ & $\cdots$ & $\cdots$ & $\cdots$ & $\cdots$ & $\cdots$ & $\cdots$ & $\cdots$ & $\cdots$ & $\cdots$ \\
\enddata
\tablecomments{Table \ref{tab:1} is published in its entirety in the machine-readable format. A portion is shown here for guidance regarding its form and content.}
\tablenotetext{a}{R.A. and Decl. are listed in J2000. Proper-motion correction should be applied when searching for the sources in ZTF data. For the median proper motion of the sources in Table~\ref{tab:1}, neglecting proper motion would lead to a typical positional offset of about $4^{\prime\prime}$ over $\sim$20 yr.}
\tablenotetext{b}{The ``Reliability" column indicates whether the period is reliable (``R") or candidate (``C").}
\tablenotetext{c}{The ``Multi." column is blank when the period of the target is derived from only one adopted band, ``Y" when multiple bands show periodicity, and ``Y$^{\star}$" when at least one additional band yields either a consistent period or an integer-day alias difference from the adopted value.}
\tablerefs{
(1) \citet{2013ApJ...779..101H};
(2) \citet{2019MNRAS.489..437D};
(3) \citet{2016ApJ...821...93N};
(4) \citet{2023AJ....165..265M};
(5) \citet{2023MNRAS.521..952M};
(6) \citet{2022MNRAS.513.2615M};
(7) \citet{\detokenize{2025A&A...695A.182M}};
(8) \citet{2021ApJ...916...77P};
(9) \citet{\detokenize{2021A&A...650A.138S}};
(10) \citet{2024MNRAS.527.8290P};
(11) \citet{2018AJ....155..196R};
(12) \citet{2011ApJ...727...56I};
(13) \citet{2006MNRAS.367..407R};
(14) \citet{2008ApJ...684..644H};
(15) \citet{2018ApJ...858...55P};
(16) \citet{2020AJ....159...60G};
(17) \citet{2017MNRAS.472.2297M};
(18) \citet{2015ApJ...799..154M};
(19) \citet{1996ASPC..109..615M};
(20) \citet{2018ApJS..237...25K};
(21) \citet{\detokenize{2024A&A...687A..95M}};
(22) \citet{\detokenize{2013A&A...554A.113H}};
(23) \citet{2004MNRAS.354..378K};
(24) \citet{2024ApJ...973..106F}
}
\end{deluxetable*}

In the ZTF/IRSA archive, photometric time series are cataloged by internal product ID, so the same sky position may yield multiple product-ID light curves. Different IDs can correspond to different bands (1/2/3 for $g/r/i$; in the 4th digit for 15-digit IDs and the 5th digit for 16-digit IDs) and different observing programs. In particular, 16-digit IDs generally correspond to private observations rather than the main public survey, so their cadence patterns and temporal baselines may differ. We therefore first analyzed all product-ID light curves independently as a preselection step, and then used same-filter spliced light curves to derive the final adopted results.

To reduce contamination from low-quality data, we applied several filtering steps to the photometric points before period analysis. First, we required \texttt{catflag} = 0 and \texttt{magerr} $>$ 0. We then performed a linear fit to the variations of R.A. and Decl. with MJD, and computed the expected position at the observation epoch for each photometry point. We excluded measurements for which the angular distance between the observed and predicted coordinates exceeded 1$^{\prime\prime}$. We did not rely on literature proper motions, as in some cases the actual ZTF coordinate changes slightly deviated from the published values. If more than 20 data points remained after this filtering, the light curve was retained for further analysis. Applying the above criteria leaves 1492 sources and reduces the number of light curves from the initial 3565 to 3309.

We used the \texttt{LombScargle} method from the \texttt{astropy.timeseries} module to compute Lomb-Scargle periodograms and obtain power spectral density. An example is presented in Figure \ref{fig:3}. Given that the rotation periods of brown dwarfs are typically no shorter than a few hours, we restricted the search to frequencies below 24 day$^{-1}$. Within this range, we identified the strongest peak with a false-alarm probability (FAP) of less than 0.001 as the period for each light curve. Following a procedure similar to \citet{2019MNRAS.486.1907J}, we assessed whether the detected peaks correspond to aliased periods, such as the common $\sim$1-day alias. If the highest peak was determined to be an alias, we excluded the period and inspected the corresponding periodogram to determine whether additional significant peaks with FAP $<$ 0.001 could be attributed to the true period. We also visually inspected both the periodograms and light curves to remove spurious results caused by abnormal sampling or cases where the light curve shows no clear variability. The light curves were then fitted using a fourth-order Fourier series, $f = a_0+\Sigma_{i=1}^4 a_i {\rm cos}(2\pi it/P+\phi_i)$, which was used to derive parameters such as the amplitude and the coefficient of determination, $R^2$. The variability amplitude is derived algebraically from the fitted coefficients as the peak-to-peak difference of the fitted Fourier series, i.e., the maximum minus the minimum value over one full phase cycle. 

We first applied the procedure above to the cleaned single product-ID light curves as a preselection step. This first-pass analysis yielded 390 period detections from individual product-ID light curves, corresponding to 300 sources. For these 300 sources, we regrouped all available light curves by source and filter, and constructed one light curve for each source-filter combination. If multiple product-ID light curves were available in the same filter, we used the light curve associated with the preliminarily selected period as the reference, shifted the others to its median magnitude, and spliced them together. Otherwise, the single available light curve was used directly. The resulting same-filter light curves were then reanalyzed using the same procedure described above. These same-filter results are adopted as the final results in this work, and 226 sources remain in the final catalog.

Because some brown dwarfs are faint and have only a small number of photometric detections, their light curves often exhibit only subtle variability, which can compromise the reliability of some period determinations. To ensure the robustness of our results, we adopted specific criteria to identify reliable periods. For a source that is relatively bright ($<17$ mag) in the band where the period is detected, we regard a period as reliable if the light curve contains more than 60 detections, $\log(\mathrm{FAP}) < -$4, and $R^{2}>0.05$. For fainter light curves, we imposed stricter criteria, requiring more than 100 detections, $\log(\mathrm{FAP}) < -$4, and $R^{2}>0.1$.

Using these selection criteria, we found that periods derived from light curves with magnitudes near the detection limit ($\sim$20 mag) typically have large amplitudes, and many such periods lie close to aliased periods. For the purposes of this work, we treat a period derived from a light curve near the detection limit as a candidate if its frequency (in units of day$^{-1}$) differs from an integer by less than 0.05. Periods that satisfy all of the criteria above are classified as reliable. In addition, even if the numerical thresholds above are not met, a period is also considered reliable if it is independently reproduced in multiple photometric bands. All remaining periods are treated as candidate periods.

Furthermore, for some sources, the periods obtained from different bands are inconsistent. For such cases, we developed a priority-based strategy to determine the final adopted period and other light-curve parameters. (1) If only one period is measured for a source, that period is adopted. (2) If multiple periods are derived, we first give preference to those classified as reliable. (3) If multiple candidate periods still remain, we compare their FAPs and adopt the one with a significantly smaller FAP. (4) If the FAP values are comparable, we select the period from the light curve with more detections. (5) If the number of detections is also comparable, we use a band-based priority: since ultracool dwarfs are brighter at longer optical wavelengths, we prioritize $i$-band periods, followed by $r$-band and then $g$-band.

\section{Results} \label{sec:5}
\subsection{Periodic Ultracool Variables Catalog} \label{sec:5.1}
The catalog of periodic ultracool variables is presented in Table \ref{tab:1} and consists of a total of 226 sources. The table includes the shortname used in the Gaia UCD Survey collaboration\footnote{https://gucds.inaf.it}, the object name in the UltracoolSheet, coordinates (R.A. and Decl. in J2000), the ZTF band in which the final adopted period was found, the average magnitude of the light curve in that band, period, logarithmic value of FAP, peak-to-peak amplitude, $R^2$ value of the light-curve fit, number of detections (after excluding low-quality photometry), spectral type, the period from the literature with its reference, the reliability of the ZTF period, and whether periodicity was detected in light curves from multiple bands. 

\begin{figure*}[ht!]
\includegraphics[width=1\textwidth]{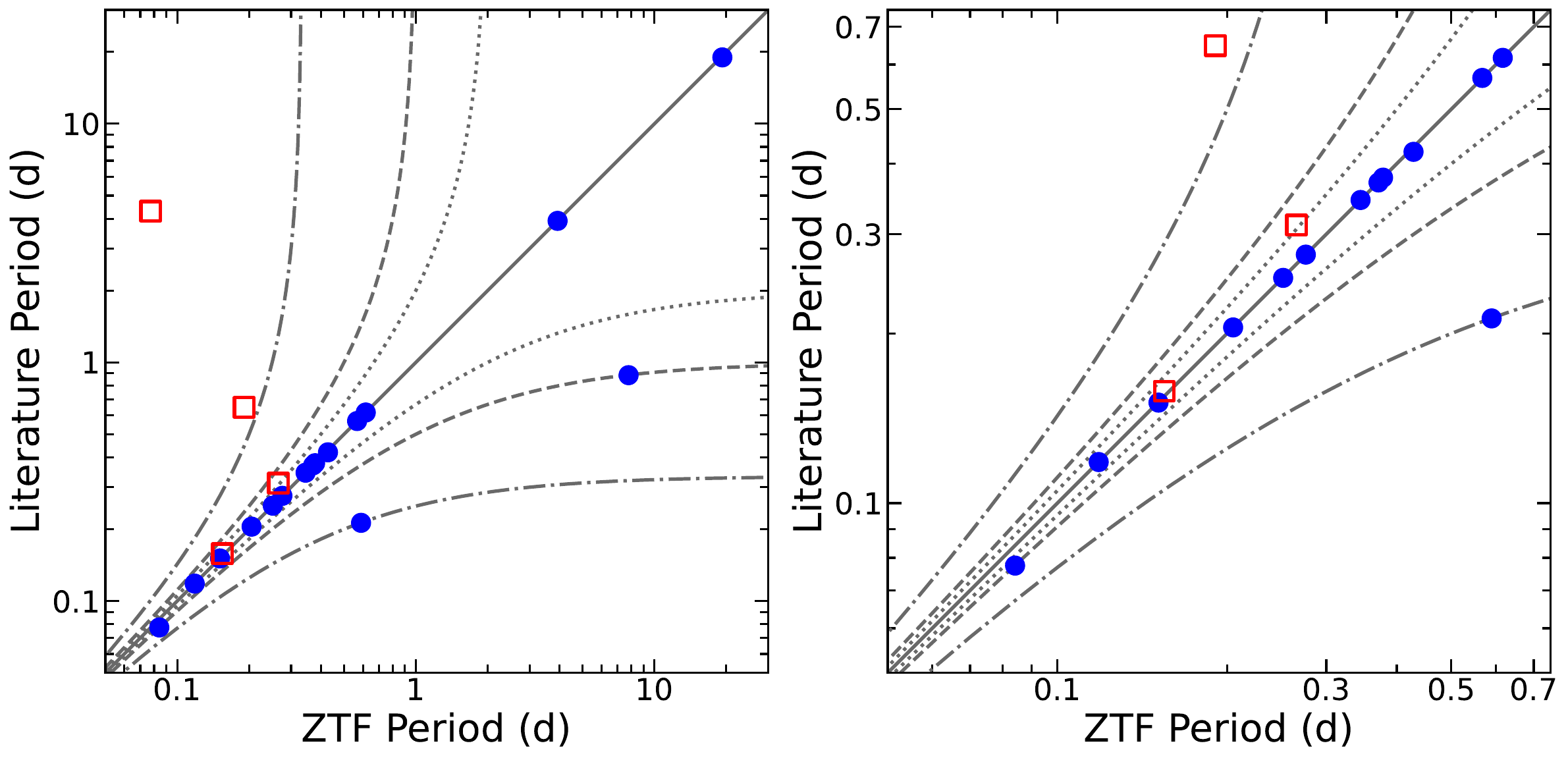}
\caption{Comparison between the literature periods and the ZTF-derived periods for our reliable results. $Left\ panel$: all sources with published periods. Filled blue circles indicate sources for which the periods are consistent or whose corresponding frequencies differ by 0.5, 1, or 3 day$^{-1}$, while red open squares denote sources whose discrepancies cannot be explained by these simple frequency offsets. The gray solid line marks equal periods, while the dotted, dashed, and dash-dotted lines represent frequency differences of 0.5, 1, and 3 day$^{-1}$, respectively. $Right\ panel$: the same comparison as in the left panel, but zoomed in to the range 0.05--0.75 days on both axes, clearly showing the slight deviation of the gray curves from the square diagonals. \label{fig:4}}
\end{figure*}

Results classified as reliable according to the criteria in Section \ref{sec:4} are marked with ``R'', while candidate periods are marked with ``C''. Reliable sources are listed at the top of Table \ref{tab:1}. Among the catalog sources, 32 are classified as reliable and 194 as candidates. For all sources with reliable adopted periods, the corresponding Figure 2- and Figure 3-style diagnostics are provided as online figure sets\footnote{\url{https://doi.org/10.5281/zenodo.20004898}}. The reliable periods are all from M-type stars, while candidate periods are more frequent for L-type stars. This spectral-type distribution reflects the selection biases of ZTF in detecting periods of ultracool dwarfs.

For some sources, periodicity is detected in more than one band. We therefore include a ``Multi.'' column in Table \ref{tab:1}. Sources with periodicity detected in only one band are left blank, while those marked with ``Y'' show periodic signals in multiple bands but with differing periods, and those marked with ``Y*'' indicate that at least one additional band yields a period consistent with, or showing an integer-day alias difference from, the adopted value. In total, 26 sources exhibit multi-band detections, including 14 marked as ``Y*'' and 12 as ``Y'', corresponding to 15 reliable and 11 candidate objects. The band-level results for these 26 sources are presented in Table \ref{tab:a1}. Most reliable multi-band sources are marked as ``Y*'', whereas candidate multi-band sources are more often marked as ``Y'', indicating that agreement between different bands is much more common among the reliable sample and further supports the robustness of those adopted periods.

\subsection{Period Accuracy} \label{sec:5.2}
Since brown dwarfs generally have short rotation periods, and ZTF provides sufficiently long temporal coverage, the primary limitation in period determination arises from the sampling cadence. For ground-based time-domain surveys, the daily cadence imposes a significant challenge to period reliability. Daily aliases are a well-known issue arising from the sampling behavior in time-series data \citep{2018ApJS..236...16V}. In contrast, most current variability studies of brown dwarfs rely on observations with shorter baselines but much higher cadence sampling, which can serve as an independent check on the reliability of our period measurements.

We searched the published literature for counterparts to our sample in order to assess the accuracy of our derived periods. The literature we consulted includes results from a variety of observations at various wavelengths, such as time-domain surveys like K2 \citep[e.g.][]{2018AJ....155..196R, 2018ApJ...858...55P}, the Transiting Exoplanet Survey Satellite \citep[TESS, e.g.][]{2024MNRAS.527.8290P, 2021A&A...650A.138S} and MEarth project \citep[e.g.][]{2011ApJ...727...56I, 2016ApJ...821...93N}, as well as targeted observations of individual objects \citep[e.g.][]{2008ApJ...684..644H}. Since most of the periods we classified as candidate correspond to relatively faint sources, they are rarely reported in the literature. The reliability of these periods cannot be guaranteed. Therefore, our comparison mainly focuses on the reliable results and their consistency with literature values. 

In Figure~\ref{fig:4}, we compare our reliable results with those reported in the literature. In this work, we adopt an operational criterion that two periods are considered consistent when their corresponding frequencies differ by no more than 0.05 day$^{-1}$. Most of our results are either directly consistent with published values or can be explained by simple alias relationships. Among the 20 reliable sources with published periods, 13 are directly consistent with the literature values, while another 3 can be explained by frequency offsets of about 1 or 3 day$^{-1}$, consistent with aliasing effects arising from ground-based sampling.

There are four remaining sources whose discrepancies are not fully accounted for by the simple alias categories above: 2MASS J01294256$-$0823580, 2MASS J08352366+1029318, LSR J0510+2713, and LSPM J1314+1320. For 2MASS J01294256$-$0823580, our ZTF period is about 0.2656 days, compared with a literature value of 0.3115 days from \citet{2019MNRAS.489..437D}. This corresponds to a frequency offset of about 0.55 day$^{-1}$. The literature value was derived from TESS Sector 3 data, whose light curve spans only about 20 days, implying a single-sector frequency resolution of $\sim 1/T \approx 0.05$ day$^{-1}$. The observed offset can therefore plausibly be explained as a 0.5 day$^{-1}$ alias.

\begin{figure}[ht!]
\includegraphics[width=0.475\textwidth]{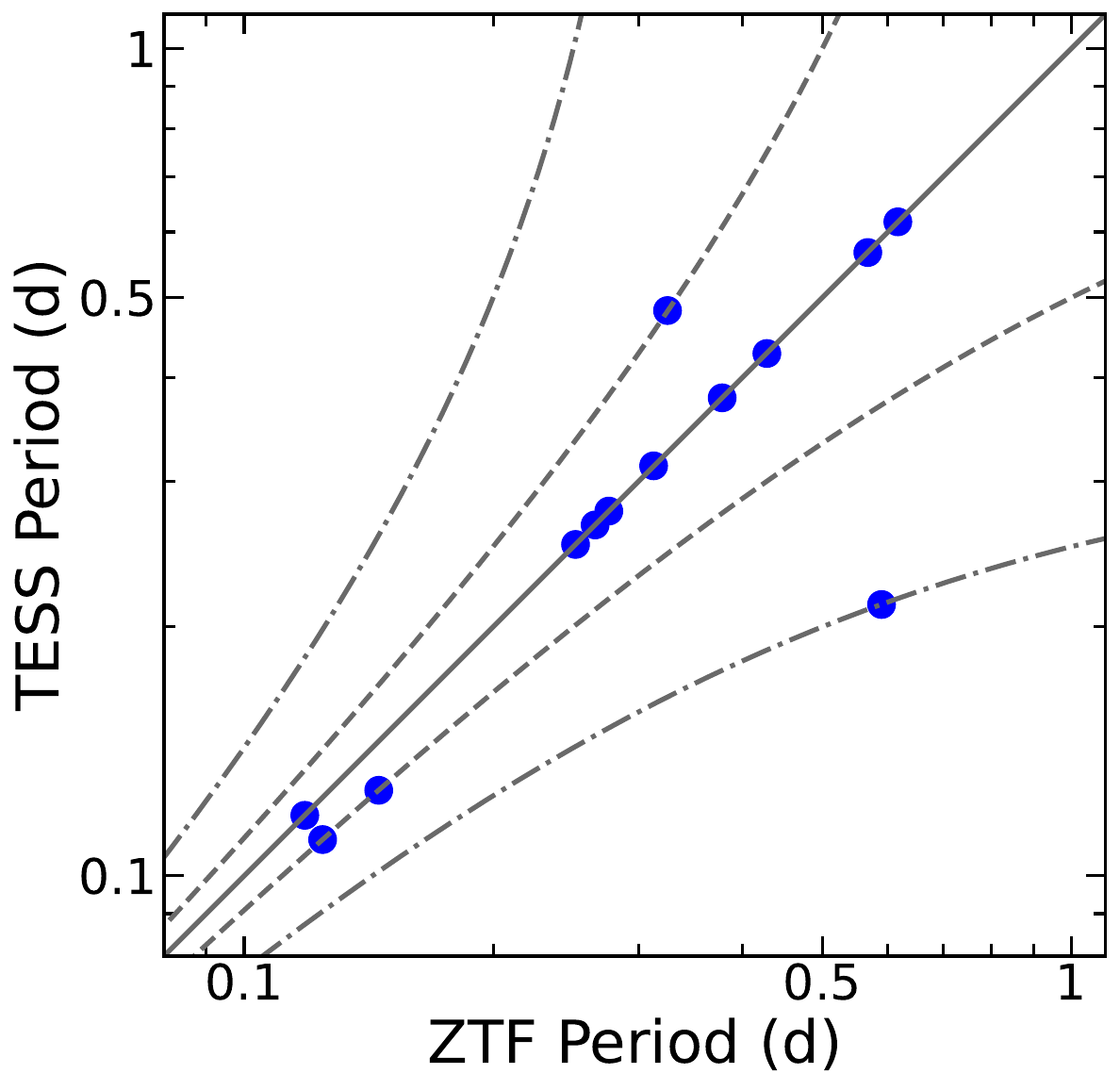}
\caption{Comparison between the ZTF-derived periods and the periods reported by Gao et al. (in prep.) from TESS photometry for the matched reliable sources. The gray solid line marks equal periods, while the dashed and dash-dotted lines indicate frequency differences of 1 and 3 day$^{-1}$, respectively. \label{fig:5}}
\end{figure}

For 2MASS J08352366+1029318, our measured period is about 0.1907 days, whereas the literature value is 0.648 days \citep{2021ApJ...916...77P}. However, in that work this source was not regarded as a significant periodic detection, and the reported value was treated as a candidate only. Similarly, for LSR J0510+2713, our ZTF period is about 0.0773 days, whereas the literature value is 4.297 days from \citet{2022MNRAS.513.2615M}. However, in that work the source was classified only as a ``B''-grade rotator rather than an ``A''-grade secure detection. As described by \citet{2022MNRAS.513.2615M}, a ``B''-grade period indicates a likely rotation signal, but not a fully secure period determination. Therefore, the discrepancies for 2MASS J08352366+1029318 and LSR J0510+2713 do not by themselves imply that our ZTF periods are incorrect.

For LSPM J1314+1320, our ZTF period is 0.1547 days ($\sim$3.7131 hr), while most literature values cluster near 3.79 hours ($\sim$0.158 d) \citep{2011ApJ...741...27M,2015ApJ...799..192W,2016ApJ...821...93N,2024A&A...690A.320D}. In our periodogram, there is indeed a peak near 3.79 hr, although it is not the highest. Notably, based on MEarth photometric monitoring, \citet{2015ApJ...799..192W} reported two periods for this source, with the secondary period measured as $3.7130 \pm 0.0002$ hr, in excellent agreement with our result. Given the extremely weak differential rotation in ultracool dwarfs \citep{2013A&A...560A...4R}, \citet{2015ApJ...799..192W} argued that the two detected periods cannot be explained by differential rotation.

In Figure \ref{fig:5}, we also compared our results with those of Gao et al. (in prep.), who derived rotation periods of ultracool dwarfs from TESS photometry. Their results include several rotation periods that previous studies did not recover in TESS data. Among 13 matched targets, 9 have periods that are essentially identical to ours, 3 show frequency offsets of $\sim$1 day$^{-1}$, and 1 shows a frequency offset of $\sim$3 day$^{-1}$ relative to the Gao et al. (in prep.) values. All four non-identical cases can be explained by simple alias relationships, further supporting the accuracy of our period determinations.

The comparison with the literature confirms that periods derived from ZTF data are generally reliable, and that the primary origin of error is indeed the one-day sampling cadence. This statement holds at least for the subset of sources we classify as reliable. However, our overall sample is dominated by results we flagged as candidate periods. Additional information on the candidate sample is provided in Appendix~\ref{sec:appendB}, including representative candidate periodograms and phase-folded light curves (Figure~\ref{fig:b1}), the candidate--literature period comparison (Figure~\ref{fig:b2}), and the distributions of spectral type and $\log(\mathrm{FAP})$ (Figure~\ref{fig:b3}). For the 16 candidates with published periods (Figure~\ref{fig:b2}), using the same frequency-based consistency criterion as for the reliable sample, two candidates agree directly with the published values, and two additional candidates are consistent with daily-alias offsets of 1 and 2~${\rm day^{-1}}$ in frequency. The remaining 12 candidates do not match the published periods within our consistency criterion. Additionally, the candidates are concentrated toward later spectral types (Figure~\ref{fig:b3}), where the currently known rotation samples are much rarer and where ZTF has more limited capability to detect and robustly measure rotation periods; they also typically show larger FAP values. We therefore treat the candidate periods as tentative and note that additional observations will be required to confirm their astrophysical origin and refine their periods, particularly for the later-type candidates, which are of special interest because they may help extend studies of rotation and atmospheric variability into the cooler ultracool-dwarf regime.

\subsection{New Variables} \label{sec:5.3}

\begin{figure}[ht!]
\includegraphics[width=0.475\textwidth]{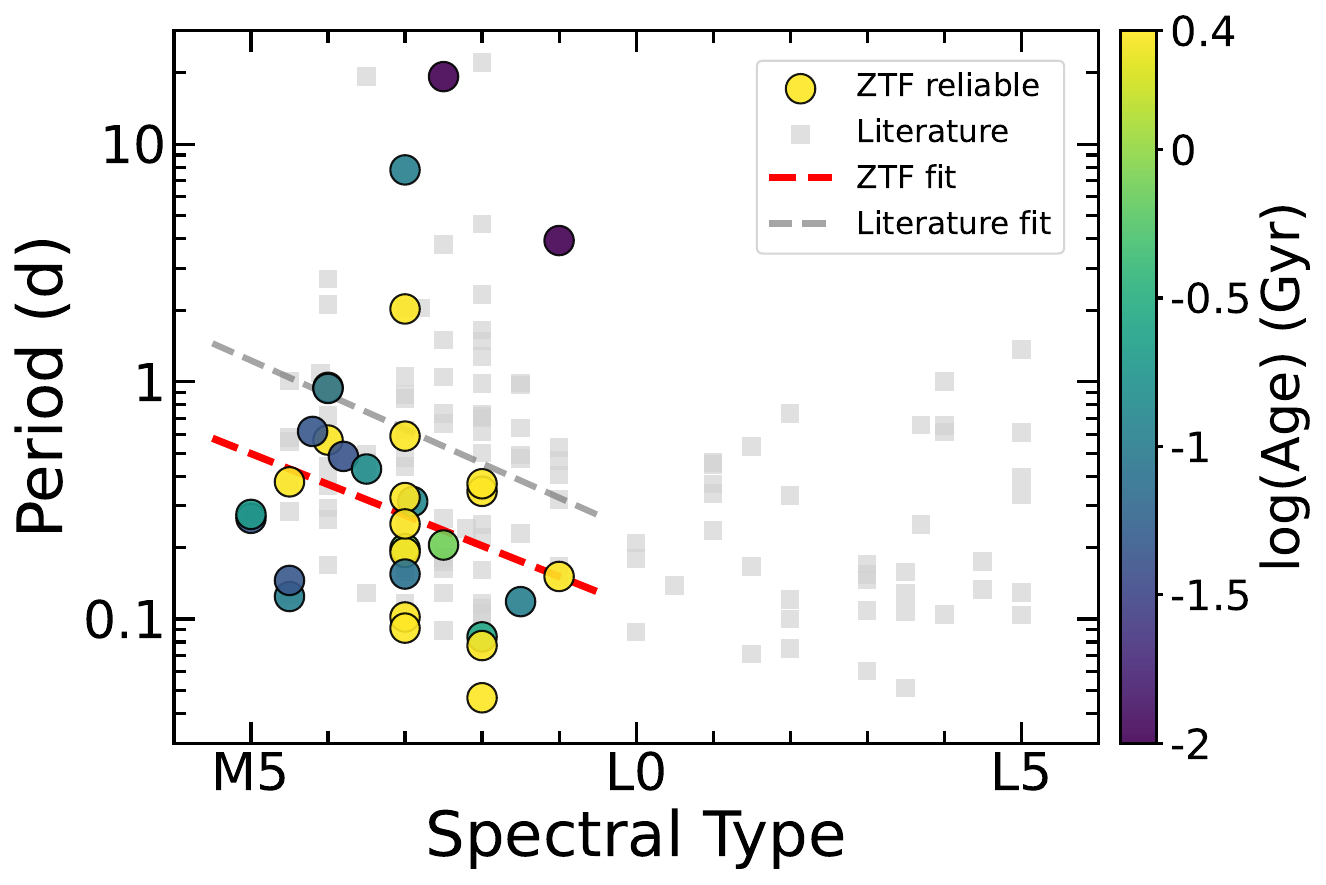}
\caption{Relation between rotation period and spectral type for the reliable sources obtained in this work. Colored circles show the ZTF reliable sample, with colors indicating ages adopted from the UltracoolSheet. Gray squares show literature rotation periods for ultracool dwarfs in the wider M5--L5 range, plotted for comparison. The red dashed line shows a linear fit to the ZTF reliable sample in $\log P$ versus spectral type, while the gray dashed line shows the corresponding fit to the M5--M9 subset of the literature comparison sample. \label{fig:6}}
\end{figure}

Of the 32 results classified as reliable, 12 have no previously reported periods and are therefore considered newly identified variables. To quantify the impact of these new periods, we compiled a literature census of published rotation periods for $\ge$M5 dwarfs with explicitly reported spectral types. The census is drawn from three sources: major ground- and space-based time-domain surveys including MEarth \citep{2016ApJ...821...93N,2018AJ....156..217N}, K2 \citep{2018AJ....155..196R, 2018ApJ...858...55P}, and \textit{TESS} \citep[e.g.][]{2023MNRAS.521..952M,2024MNRAS.527.8290P,2024ApJ...973..106F}; recent individual-object and small-sample monitoring studies \citep{2017MNRAS.472.2297M,2021A&A...651L...7M,2022MNRAS.513.3482A,2022ApJ...924...68V}; and the compiled L--Y dwarf period tables of \citet{2021AJ....161..224T}. Some individual-object and small-sample studies may not be fully represented. The census contains 196 unique objects; our 12 newly established periods increase the published $\ge$M5 period sample by $\sim$6\%. All 12 new periods fall in the M5--M9 range, which includes 91 objects in the census, corresponding to roughly twice the fractional increase in this spectral-type interval.

\begin{figure}[ht!]
\includegraphics[width=0.475\textwidth]{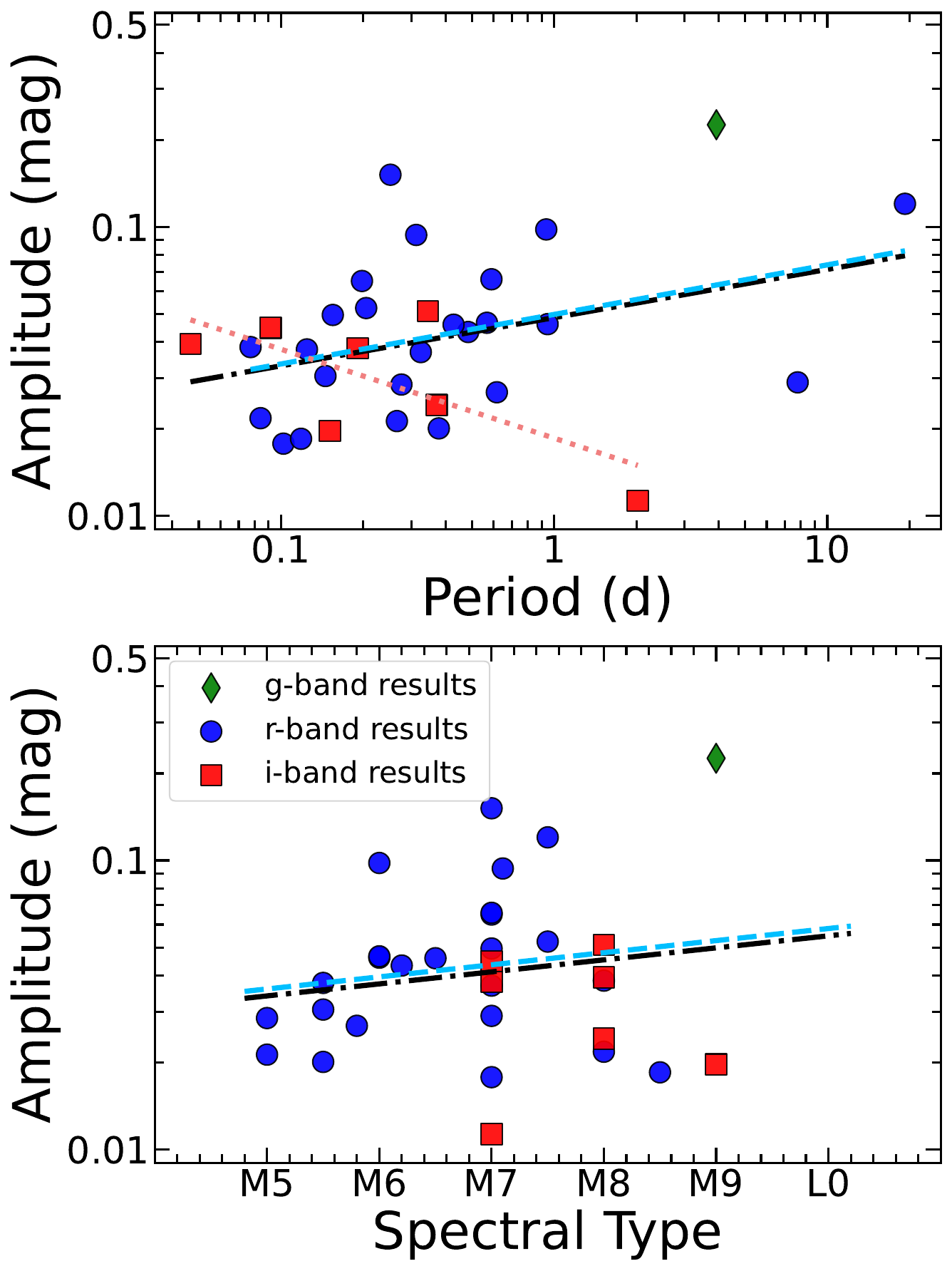}
\caption{
Peak-to-peak amplitude as a function of rotation period (top) and spectral type (bottom) for sources with reliable period measurements. Different colors indicate the bandpasses from which the parameters were derived. In the top panel, the black dash-dotted line shows $\log_{10}(A)=0.167\log_{10}(P)-1.314$ for the full reliable sample (Pearson $r=0.336$), the light blue dashed line shows $\log_{10}(A)=0.171\log_{10}(P)-1.303$ for the $r$-band subsample (Pearson $r=0.376$), and the light red dotted line shows $\log_{10}(A)=-0.308\log_{10}(P)-1.731$ for the $i$-band subsample (Pearson $r=-0.687$). In the bottom panel, the black dash-dotted line shows $\log_{10}(A)=0.042S-1.676$ for the full reliable sample ($r=0.155$). The light blue dashed line shows the corresponding fit for the $r$-band subsample, $\log_{10}(A)=0.042S-1.654$ ($r=0.156$). \label{fig:7}}
\end{figure}

\section{Discussion} \label{sec:6}

The rotation periods of M dwarfs are correlated with their masses. Statistical studies have shown that, within the long-period regime \citep[on the order of tens of days,][]{2016ApJ...821...93N} of M dwarfs, lower-mass objects tend to have longer rotation periods and therefore rotate more slowly. In the short-period regime \citep[with timescales of hours,][]{2016ApJ...821...93N}, by contrast, the trend is reversed, lower-mass objects tend to have shorter periods and correspondingly faster rotation \citep{2011ApJ...727...56I, 2013MNRAS.432.1203M, 2016ApJ...821...93N}. This pattern is also reflected in projected rotational velocity $v \sin i$ measurements of brown dwarfs \citep{2003ApJ...583..451M, 2006ApJ...647.1405Z}. It has also been reported that the color–period relation mirrors the mass–period relation, exhibiting similar differences between the long- and short-period populations \citep[Figure 15 therein]{2016ApJ...821...93N}. 

\begin{figure}[ht!]
\includegraphics[width=0.475\textwidth]{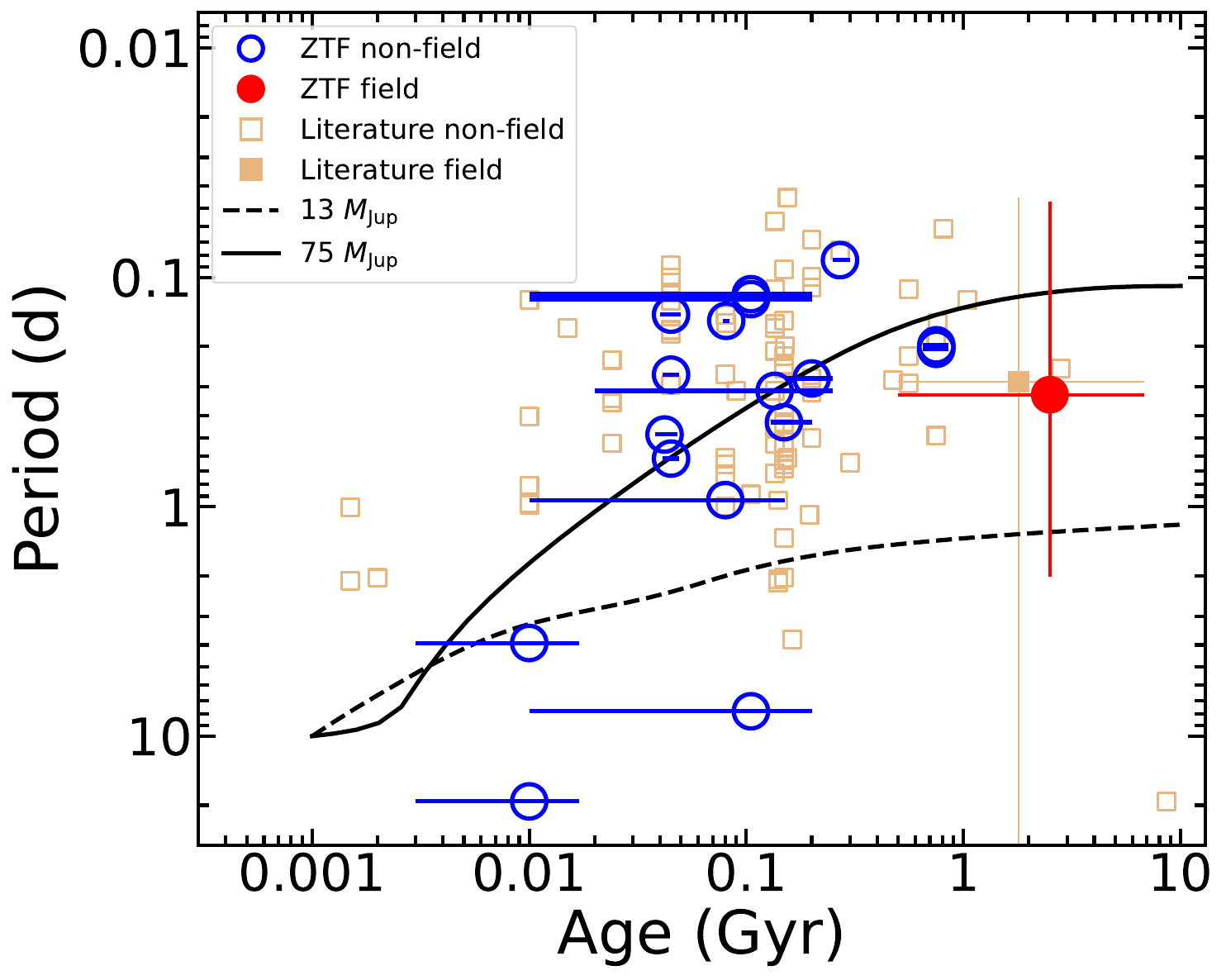}
\caption{Relation between age and rotation period for the reliable sources in this work. Ages and their uncertainties are adopted from the UltracoolSheet. Blue open circles denote ZTF non-field sources, while the red filled circle represents the ZTF field population, plotted using the median period and the full period range of objects classified as ``Field'' in the UltracoolSheet. Literature ultracool dwarfs are overplotted for comparison: open squares denote non-field objects and the filled square denotes the literature field population, summarized in the same way. The literature field point is shifted slightly along the x-axis for visual clarity. Horizontal error bars indicate age uncertainties; for some sources with similar ages and periods, their points and error bars may overlap. The black solid and dashed lines show angular-momentum-conservation tracks for an initial period of 10 days at 75 and 13~$M_{\rm Jup}$, respectively, based on the radius evolution model of \citet{2023A&A...671A.119C}. \label{fig:8}}
\end{figure}

In Figure~\ref{fig:6}, we examine how the rotation periods of our reliable sources vary with spectral type. The ZTF reliable sample alone shows only a weak tendency for shorter periods at later spectral types. A linear fit in $\log P$ versus spectral type gives a negative slope of $-0.129$, but the correlation is weak (Pearson $r=-0.293$). To place this result in a broader context, we also overplot literature rotation periods for ultracool dwarfs spanning the wider M5--L5 range as gray points, compiled from published studies of ultracool-dwarf rotation \citep{2016ApJ...821...93N,2017ApJ...840...83M,2017MNRAS.472.2297M,2018AJ....156..217N,2021AJ....161..224T,2021A&A...651L...7M,2022ApJ...924...68V,2023MNRAS.521..952M,2024ApJ...973..106F,2024MNRAS.527.8290P,2025AJ....170..168L}. Because the ZTF reliable sample spans the M5--M9 range, and the literature sample does not exhibit an equally clear monotonic decline of period with spectral type beyond L0, we restrict the comparison fit to the M5--M9 subset of the literature sample. This fit also yields a negative slope ($-0.144$) and a Pearson correlation coefficient of $r=-0.233$, both close to the corresponding values derived from our ZTF reliable sample, indicating that the tentative trend seen in the ZTF sample is qualitatively consistent with the literature trend over the same spectral-type range.

The amplitudes of variability in brown dwarfs also correlate with spectral type. Objects at the L/T transition typically show the largest amplitudes \citep[e.g.][]{2014ApJ...797..120R}. Unfortunately, ZTF cannot detect sources of such late types, and our sample is therefore dominated by M dwarfs. For M dwarfs, the variability amplitude is correlated with period, the shorter the period (i.e., the faster the rotation), the larger the amplitude \citep{2014ApJS..211...24M, 2016ApJ...821...93N}. Figure \ref{fig:7} shows the relations of peak-to-peak amplitude with rotation period (top) and spectral type (bottom), for our reliable results. Because parameters for different sources come from different bands, the measured amplitudes may differ systematically. Therefore, we plot the $g$-, $r$-, and $i$-band results with green diamond, blue circles, and red squares, respectively. To quantify the amplitude--period relation, we fit the data with $\log_{10}(A)=k\,\log_{10}(P)+b$, where $A$ is the peak-to-peak amplitude and $P$ is the rotation period. The full reliable sample shows only a weak positive correlation (Pearson $r=0.336$), and the $r$-band subsample shows a similarly weak positive correlation (Pearson $r=0.376$). Since most reliable measurements are in the $r$ band, the full-sample trend is largely dominated by the $r$-band results. However, the $i$-band subsample shows the opposite tendency (Pearson $r=-0.687$). Therefore, our sample does not show a clear and band-independent amplitude--period relation.

The relation between variability amplitude and spectral type is also examined. We also linearly fit the amplitude--spectral-type relation with $\log_{10}(A)=k\,S+b$ for both the full reliable sample and the $r$-band subsample, where $A$ is the peak-to-peak amplitude and $S$ is the spectral-type index (M5=5, L0=10). Both fits show only a very weak positive correlation (full: $r=0.155$, $r$-band only: $r=0.156$). The two fits are nearly identical, likely because most measurements are from the $r$ band.

Brown dwarfs contract continuously during their evolution, and under the assumption of angular momentum conservation their rotation rates increase and their rotation periods decrease \citep{2023A&A...671A.119C}. In Figure~\ref{fig:8}, we present the relation between age and rotation period for our reliable results, with ages adopted from the UltracoolSheet and compiled by \citet{2023ApJ...959...63S}. Because many sources cannot be associated with a stellar association or moving group and are therefore classified as field objects, a large fraction of ages are uniformly assigned as 2.5~Gyr based on the field-age distribution adopted by \citet{2017ApJS..231...15D}. Such age estimates carry substantial uncertainties and should therefore be interpreted with caution. For these field sources, we use the median rotation period and the full period range to represent the field population in Figure~\ref{fig:8}. To place our results in a broader context, we also overplot literature ultracool dwarfs, separating non-field and field objects in the same way.

Overall, Figure~\ref{fig:8} supports the expected spin-up of ultracool dwarfs with age, in the sense that older objects tend to have shorter rotation periods. We also compare the data with the radius-evolution tracks of \citet{2023A&A...671A.119C} under angular momentum conservation, shown for two limiting masses in Figure~\ref{fig:8}. The age-period relation of our sample is more consistent with the 75~$M_{\rm Jup}$ track than with the 13~$M_{\rm Jup}$ track, likely because our sample is dominated by M dwarfs, which on average have relatively higher masses.

\section{Conclusion} \label{sec:7}

We have conducted a comprehensive photometric survey of ultracool dwarfs using ZTF optical light curves. ZTF's multi-year baseline provides unique insight into long-term atmospheric stability that complement high-cadence space-based surveys. However, its optical sensitivity fundamentally limits studies to brighter, earlier-type objects. Through precise astrometric corrections to account for high proper motions, an initial single-product-ID search, same-filter splicing, and Lomb-Scargle periodogram analysis, we compiled a final catalog of 226 periodic ultracool variables. Of these catalog sources, 32 detections were classified as reliable and 194 as candidates. The reliable detections are confined to M dwarfs, a direct consequence of ZTF's optical sensitivity limits, which favor nearby brighter, earlier-type objects. 12 of these reliable variables are new discoveries, expanding the census of rotationally variable ultracool dwarfs. Among the reliable sources with literature counterparts, most are either directly consistent with published values or can be explained by simple alias relationships, supporting the robustness of our methodology despite the inherent challenge of daily aliasing in ground-based surveys. Our catalog establishes ZTF as a powerful resource for long-term rotational monitoring of nearby substellar objects.

Physical analysis reveals trends in our sample. Field dwarfs older than 100 Myr show decreasing rotation periods toward later spectral types, indicating faster rotation in late-M dwarfs compared to their mid-M counterparts in the short-period regime. We also find no clear band-independent relation between rotation period and variability amplitude, while the amplitude--spectral-type relation shows at most a very weak positive trend in our reliable sample. Age--period relationships align closely with evolutionary models assuming angular momentum conservation in the higher mass regime, consistent with the predominantly stellar masses in our M-dwarf sample.

Beyond the robust detections, our survey has identified a substantial number of promising periodic candidates that exhibit periodogram peaks but fall short of our strict reliability criteria due to insufficient sampling or signal-to-noise ratio. Follow-up monitoring with higher cadence, such as that achievable with the Large Synoptic Survey Telescope (LSST), or extended temporal baseline observations will be essential to confirm their periodic nature, thereby significantly expanding the census of rotationally variable ultracool dwarfs. Furthermore, the ZTF dataset holds considerable potential for discovering new late-type variables independently of existing catalogs. By applying photometric and astrometric selection techniques, such as multi-band color cuts and proper motion screening, directly to the ZTF database, future work could systematically identify previously unrecognized periodic M and L dwarfs. This approach would provide a more complete picture of rotational evolution and atmospheric heterogeneity across the stellar–substellar boundary.

\begin{acknowledgments}
We thank the anonymous referee for the helpful comments. This work is supported by the National Natural Science Foundation of China (NSFC) through the projects 12588202, 12373028, 12322306, 12173047, and 12133002. S. W. and X. C. acknowledge support from the Youth Innovation Promotion Association of the CAS with Nos. 2023065 and 2023055. This work is based on observations obtained with the Samuel Oschin Telescope 48-inch and the 60-inch Telescope at the Palomar Observatory as part of the Zwicky Transient Facility project. ZTF is supported by the National Science Foundation under Grants No. AST-1440341 and AST-2034437 and a collaboration including current partners Caltech, IPAC, the Oskar Klein Center at Stockholm University, the University of Maryland, University of California, Berkeley, the University of Wisconsin at Milwaukee, University of Warwick, Ruhr University, Cornell University, Northwestern University and Drexel University. Operations are conducted by COO, IPAC, and UW.
\end{acknowledgments}



\appendix

\section{Period Comparison among Different ZTF Bands} \label{sec:appendA}
\restartappendixnumbering
\begin{deluxetable}{llcclcrccc}\label{tab:a1}
\tabletypesize{\scriptsize}
\tablecaption{Comparison of Periods Derived from Different ZTF Bands}
\tablehead{\colhead{Shortname} & \colhead{Name} & \colhead{Band} & \colhead{Magnitude} & \colhead{N} & \colhead{Period} & \colhead{log(FAP)} & \colhead{Amplitude} & \colhead{R$^2$} & \colhead{Remark} \\ 
\colhead{} & \colhead{} & \colhead{} & \colhead{(mag)} & \colhead{} & \colhead{(days)} & \colhead{} & \colhead{(mag)} & \colhead{} & \colhead{} } 
\startdata
 & WISEA J000538.83+020951.2 & r & 16.731 & 430 & 0.9461404 & -23.3 & 0.046 & 0.254 & A \\
 &  & g & 18.464 & 349 & 0.3268514 & -5.5 & 0.058 & 0.114 & I \\
 & PSO J012.7+007.0 & r & 19.130 & 481 & 0.3125428 & -12.1 & 0.094 & 0.142 & A \\
 &  & i & 16.455 & 103 & 0.3125446 & -11.9 & 0.059 & 0.595 & M \\
J0217+3526 & LP 245-10 & r & 14.912 & 490 & 0.2758886 & -13.1 & 0.029 & 0.160 & A \\
 &  & g & 16.690 & 411 & 0.2758994 & -6.5 & 0.031 & 0.108 & M \\
J0235-0711 & DENIS J023549.5-071121 & r & 17.716 & 625 & 0.3248738 & -11.5 & 0.037 & 0.111 & A \\
 &  & i & 15.287 & 68 & 0.3248567 & -3.6 & 0.027 & 0.377 & M \\
J0311+0106 & 2MASS J03111547+0106307 & r & 15.278 & 418 & 0.1244036 & -17.0 & 0.038 & 0.219 & A \\
 &  & g & 17.008 & 364 & 0.1106063 & -10.6 & 0.065 & 0.173 & I \\
J0330+2405 & 2MASS J03300506+2405281 & r & 17.617 & 929 & 0.1978335 & -33.5 & 0.065 & 0.192 & A \\
 &  & i & 15.212 & 74 & 0.3282928 & -6.6 & 0.040 & 0.497 & I \\
 &  & g & 19.395 & 458 & 0.5651446 & -3.2 & 0.067 & 0.054 & S \\
J0357+4107 & 2MASS J03571999+4107426 & r & 15.986 & 1844 & 0.5672745 & -100.6 & 0.047 & 0.246 & A \\
 &  & g & 17.745 & 849 & 0.5672712 & -10.7 & 0.038 & 0.088 & M \\
J0752+1612 & LP 423-31 & r & 16.041 & 580 & 7.7883021 & -34.1 & 0.029 & 0.341 & A \\
 &  & g & 17.813 & 270 & 0.4686574 & -5.3 & 0.071 & 0.153 & I \\
J0807+3213 & 2MASS J08072607+3213101 & i & 15.502 & 63 & 0.3450465 & -4.6 & 0.051 & 0.488 & A \\
 &  & r & 18.177 & 1257 & 0.3450082 & -18.8 & 0.049 & 0.086 & M \\
J1314+1319B & LSPM J1314+1320 & r & 14.737 & 402 & 0.1547113 & -22.5 & 0.050 & 0.266 & A \\
 &  & i & 12.447 & 100 & 0.1547109 & -4.6 & 0.035 & 0.334 & M \\
 &  & g & 16.492 & 258 & 0.1547114 & -3.4 & 0.028 & 0.139 & M \\
J1438+5559 & 2MASS J14384542+5559134 & i & 16.006 & 790 & 2.0225841 & -7.8 & 0.011 & 0.067 & A \\
 &  & r & 18.406 & 2133 & 2.0238634 & -14.1 & 0.020 & 0.045 & M \\
J1640+6736 & G 240-23 & r & 14.663 & 685 & 0.3782275 & -6.5 & 0.020 & 0.072 & A \\
 &  & g & 16.445 & 610 & 0.5741471 & -4.5 & 0.033 & 0.062 & F \\
J2307+0520 & WISEA J230743.63+052037.3 & i & 17.698 & 141 & 0.0916088 & -5.3 & 0.045 & 0.249 & A \\
 &  & r & 20.334 & 388 & 0.0418612 & -3.3 & 0.130 & 0.069 & L \\
J2338+0624 & LP 583-21 & r & 16.309 & 69 & 0.2514961 & -13.3 & 0.151 & 0.761 & A \\
 &  & i & 14.143 & 36 & 0.2514736 & -6.5 & 0.111 & 0.721 & M \\
 &  & g & 18.029 & 49 & 0.0794266 & -4.8 & 0.193 & 0.553 & F \\
J2353-1844 & 2MASS J23532556-1844402 & r & 15.762 & 482 & 0.6167348 & -12.8 & 0.027 & 0.169 & A \\
 &  & g & 17.421 & 532 & 0.3810549 & -7.9 & 0.044 & 0.130 & I \\
J0320+1854 & LP 412-31 & g & 19.881 & 108 & 0.1427245 & -5.6 & 0.425 & 0.261 & A \\
 &  & i & 15.106 & 38 & 1.5969935 & -4.2 & 0.066 & 0.592 & S \\
J0357+1529 & SIMP J03570493+1529270 & i & 18.982 & 23 & 0.0486741 & -4.3 & 0.217 & 0.832 & A \\
 &  & r & 21.251 & 43 & 0.1385560 & -3.0 & 0.565 & 0.511 & F \\
J0735+4108 & SDSS J073519.59+410850.4 & i & 19.174 & 71 & 0.1366597 & -3.4 & 0.119 & 0.368 & A \\
 &  & r & 21.459 & 32 & 0.4793582 & -3.2 & 0.689 & 0.667 & S \\
$\cdots$ & $\cdots$ & $\cdots$ & $\cdots$ & $\cdots$ & $\cdots$ & $\cdots$ & $\cdots$ & $\cdots$ & $\cdots$ \\
\enddata

\tablecomments{
Table \ref{tab:a1} is published in its entirety in the machine-readable format. A portion is shown here for guidance regarding its form and content.\\
Codes in the last column indicate the reasons for differences between band-level periods:
A – adopted (period used in Table 1);
M – match (identical to the adopted period within the tolerance);
I – alias (consistent with an integer day$^{-1}$ frequency offset from the adopted value);
S – sparse (light curve with significantly fewer epochs);
F – fainter (the mean flux is significantly lower than that of the adopted light curve);
L – limit (light curve near the ZTF detection limit).\\
Periods marked as S, F, or L are considered less reliable and are therefore not recommended for adoption.}
\end{deluxetable}
\onecolumngrid

For targets with periodicity detected in more than one ZTF band, we list all the results in Table \ref{tab:a1}. Columns include the target name, band, mean magnitude in that band, number of data points, period, log(FAP), peak-to-peak amplitude, and R$^2$. Results from different bands for the same source are listed in adjacent rows. The last column (“Remark”) indicates whether the period is the adopted value (``A") in Table \ref{tab:1}; matches the adopted period (``M"), shows an integer-day alias difference caused by sampling (``I"), or differs from the adopted value with possible reasons (``S", ``F", and ``L", see the comments of Table \ref{tab:a1} for more details). The latter three codes (``S'', ``F'', and ``L'') indicate that the corresponding band-level periods may be less reliable and are therefore not recommended for adoption. The full version of Table \ref{tab:a1} is available in machine-readable format.

\section{Candidate Periods}\label{sec:appendB}

\begin{figure}[t!]
\includegraphics[width=1\textwidth]{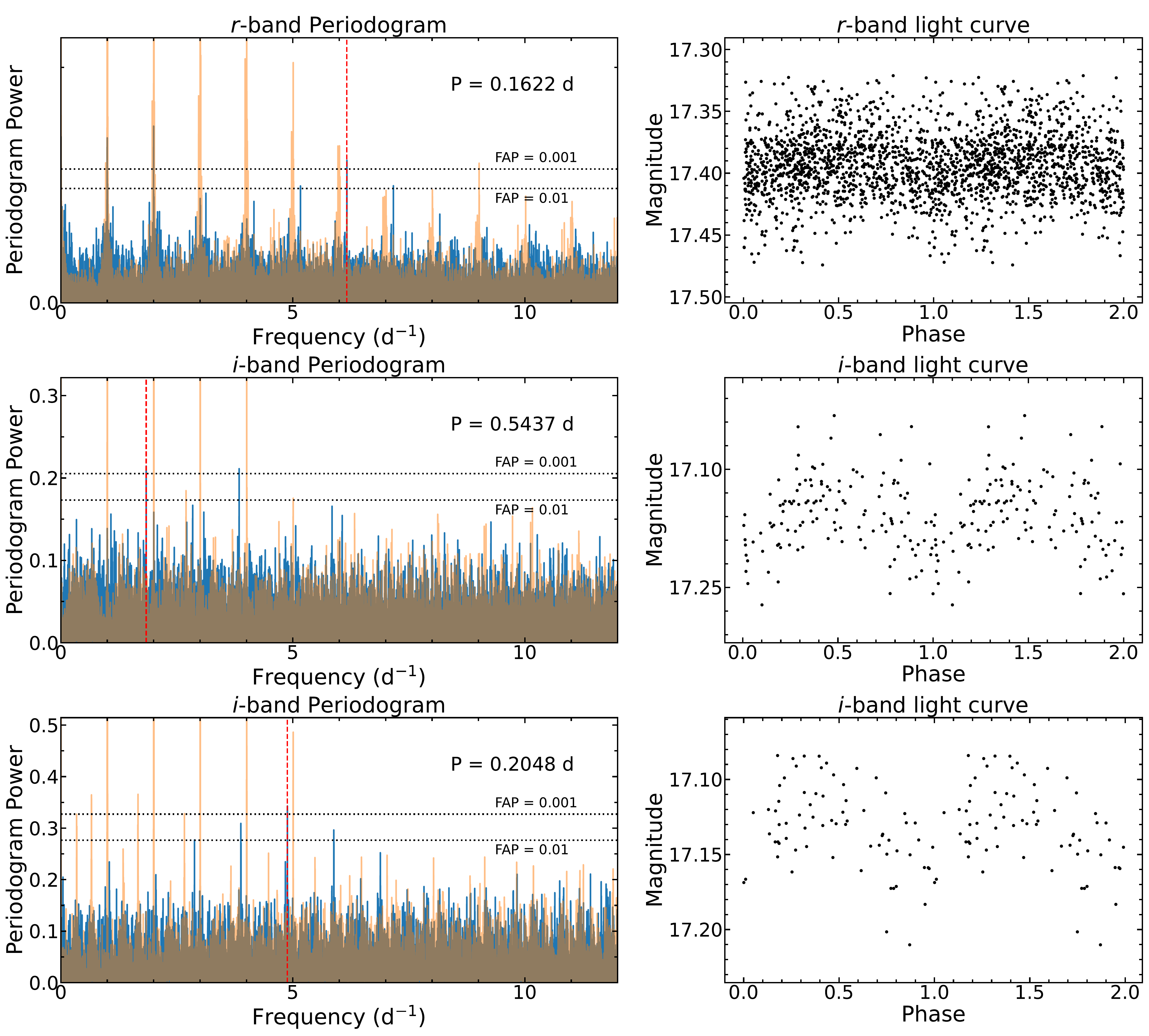}
\caption{Same as Figure~\ref{fig:3}, but showing three candidate examples: LP 44-162 (top; consistent with the literature), 2MASS J14122268+2354108 (middle; a 2~day$^{-1}$ alias offset in frequency relative to the literature), and TVLM 831-161058 (bottom; no published period available for comparison).\label{fig:b1}}
\end{figure}

\begin{figure}[ht!]
\includegraphics[width=1\textwidth]{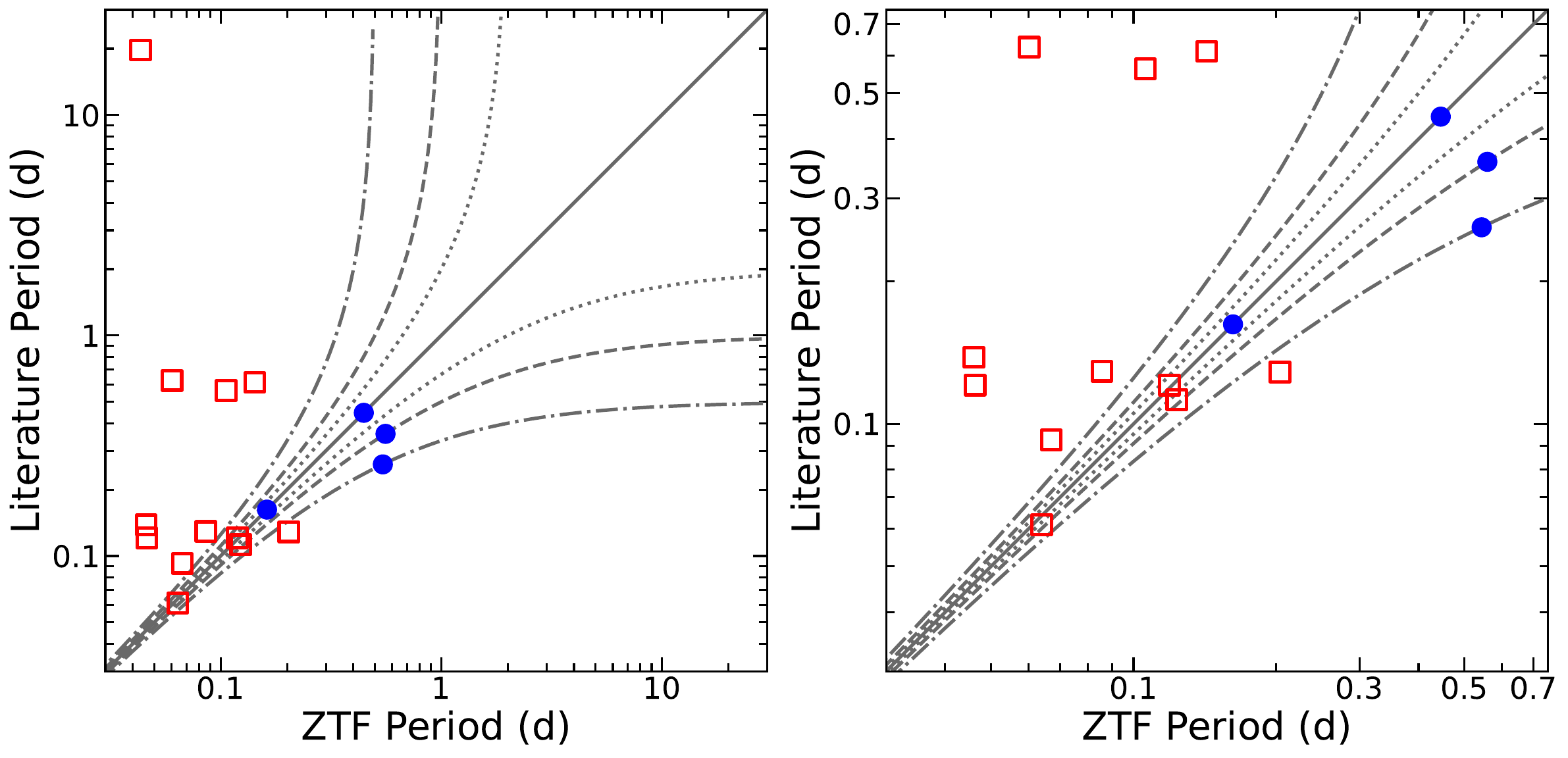}
\caption{Same as Figure~\ref{fig:4}, but for the candidate periods with available literature counterparts. Here the gray dotted, dashed, and dash-dotted lines indicate frequency offsets of 0.5, 1, and 2~day$^{-1}$, respectively. \label{fig:b2}}
\end{figure}

\begin{figure}[ht!]
\includegraphics[width=1\textwidth]{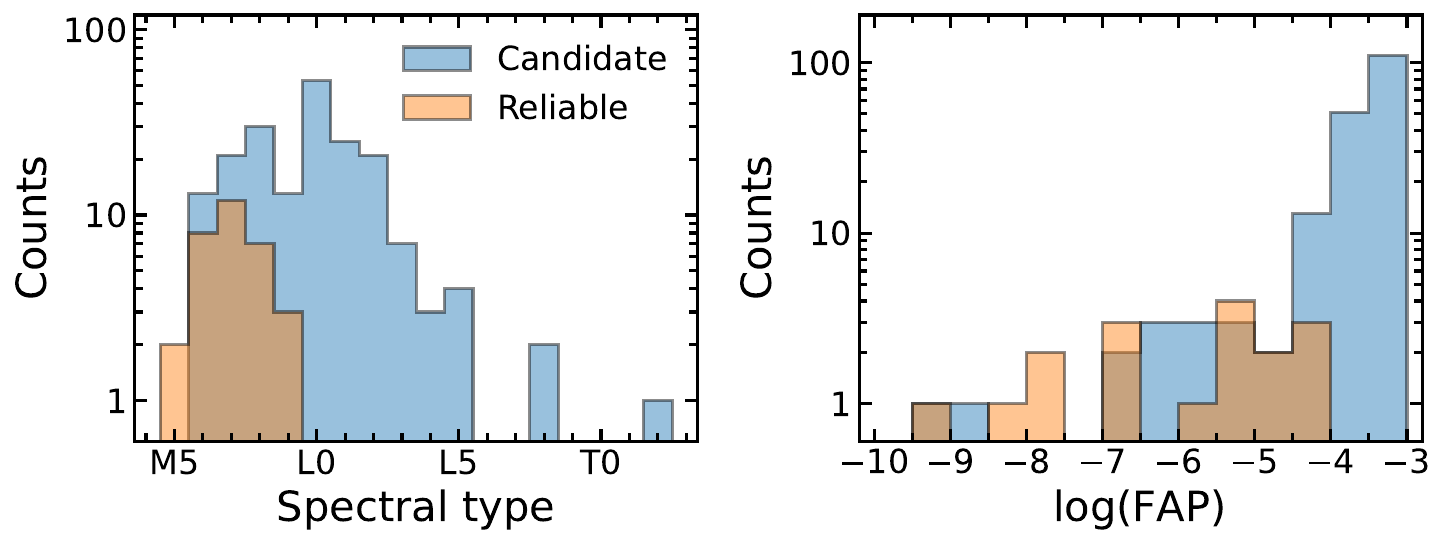}
\caption{Distributions of spectral type (left) and $\log(\mathrm{FAP})$ (right) for the reliable and candidate samples. For the $\log(\mathrm{FAP})$ panel we show the range $-10\le \log(\mathrm{FAP}) \le -3$, where most candidates lie. \label{fig:b3}}
\end{figure}

To complement the main analysis focused on the reliable sample, we provide additional information on the candidate periods. Figure~\ref{fig:b1} presents three representative candidate examples, illustrating cases where the ZTF period is consistent with the literature, differs by a 2~day$^{-1}$ alias offset in frequency, or has no published period available for comparison. Compared with the reliable examples in Figure~\ref{fig:3}, the candidates typically show lower periodogram peak powers (and hence larger FAP values) and less clearly defined phase-folded modulations. Figure~\ref{fig:b2} compares the ZTF-derived candidate periods with literature values for the 16 candidates with available counterparts, adopting the same frequency-based consistency criterion as for the reliable sample in Section~\ref{sec:5.2}. Only 4 of these candidates are consistent with the published values under the same frequency-based criterion used for the reliable sample, the other 12 still show periodogram peaks in the ZTF light curves, but their inferred periods are not yet reliable enough to be treated as confirmed rotation periods. These candidates therefore require further validation with additional ZTF coverage and/or independent higher-cadence observations. Figure~\ref{fig:b3} further summarizes the candidate population by comparing the distributions of spectral type and $\log(\mathrm{FAP})$ between the candidate and reliable samples. The candidates are concentrated toward later spectral types and typically have less significant peaks (larger FAP), consistent with their lower reliability under our selection criteria. Additional observations will be particularly valuable for these later-type candidates, because currently known rotation samples in this spectral-type regime are much rarer, and confirming their variability would help extend studies of rotation and atmospheric variability to cooler ultracool dwarfs.

\bibliography{brown_dwarf_ztf_revise2}{}
\bibliographystyle{aasjournalv7}

\end{document}